\documentclass[11pt]{article}

\usepackage[margin=1in]{geometry}
\usepackage{amsmath,amssymb,amsfonts}
\usepackage{graphicx}
\usepackage{hyperref}
\usepackage{braket}
\usepackage{comment}
\usepackage{gensymb}
\usepackage{xcolor}

\pdfoutput=1

\definecolor{my_green}{rgb}{0.55, 0.71, 0.0}

\newcommand{\imagi}{\mathsf{i}}

\title{Quantum-Tunnelling Oscillators for Cognitive Modelling and Neural Computation: Foundations, Machine-Vision Realisation and Applications}

\author{
Ivan S.~Maksymov\\
Seymour Research Laboratories, Seymour, VIC 3660, Australia\\
\texttt{ivan.maksymov@gmail.com}
}

\date{\today}

\begin{document}

\maketitle

\begin{abstract}
I present a quantum-tunnelling oscillator model as a universal dynamical engine for two paradigmatic problems in quantum cognition theory—optical illusion perception and group decision making—where individuals are treated as quantum-mechanical agents whose choices shift through context-dependent transitions rather than simple probabilities. I show that, when networked together, these units form a quantum-cognitive neural system that reproduces familiar collective and perceptual phenomena while naturally accommodating counterintuitive processes that challenge classical models. Bridging ideas from quantum cognition theory and neural networks, this approach offers a compact, physically grounded way to describe how real individuals and groups think, perceive and decide.
\end{abstract}


\section{Introduction}
\setcounter{subsection}{-1}
\subsection{Personal Motivation}
This paper presents original results from a sustained effort to build a physically grounded model of individual and collective behaviour under uncertainty, risk and cognitive pressure, including pressure exerted by diffuse and often unaccountable actors. My primary motivation to write it is Ukraine, its surrounding regions and people who live there or were born there but moved abroad before the current war and in its uncertain aftermath. The cognitive environment there has shifted so profoundly that people once familiar, including family, friends and colleagues, have become difficult to recognise.

Forced to leave Ukraine well before these events, I observed the situation from Australia. That distance sharpened rather than dulled the impulse to act. My response was to turn to quantum physics, supplemented by state-of-the-art literature on quantum cognition and decision-making. This paper summarises the result:~roughly three years of work to translate that impulse into a formal, testable framework.  

\subsection{Scientific Motivation}
The study of decision-making under risk and uncertainty continues to reveal systematic irregularities. In particular, individuals often revise their choices when identical decision problems are repeated over short intervals \cite{hey1994investigating, blavatskyy2010endowment}. Such behaviour sits uneasily with classical deterministic decision theories, which assume stable preferences under invariant conditions \cite{blavatskyy2010models}. The tension is especially visible in data derived from social networks~\cite{Bai18, Khr23, Tok21, Gal24, Gal_book}, finance~\cite{Lee20, Pkhr24, Bag25, Jam26}, international relations~\cite{Tes15, Bra23, Wat24}, defence sector~\cite{Hum25, Hum25_1}, psychology~\cite{Cle13, Cle17} and visual perception~\cite{Ein04, Kor05, Bus12, Ben18, Joo20, Ara20}. It is also prominent in interactive environments such as video games~\cite{Bav12, Poh22, Li15, wang2021game, Neckerworld}, where repeated exposure amplifies variability in decision-making~\cite{bailey2013would}. 

In this paper, I advance a quantum-tunnelling oscillator model of decision-making that links established psychological phenomena with formal constructs from physics, with particular emphasis on human behaviour and perception and their modelling by means of neural networks. Building of my previous work in this field and the adjacent domains~\cite{Mak23_review}, I demonstrate that this approach captures observed personal and social behavioural regularities more effectively than conventional approaches.

\begin{figure}
 \centering
 \includegraphics[width=0.999\textwidth]{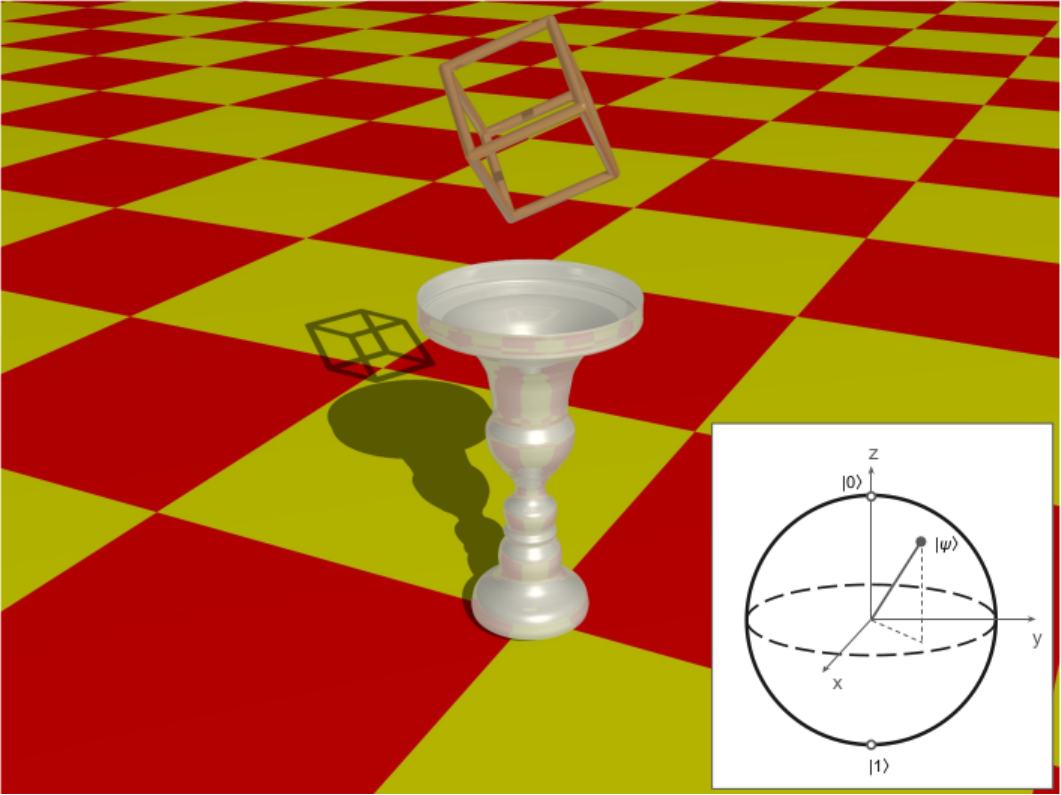}
 \caption{Conceptual illustration generated using ray-tracing software, showing the shadows cast by a metal wire cube and a silver vase. While the three-dimensional objects are perceived unambiguously, careful observation of their two-dimensional projections reveals bistable perception:~the shadow of the cube can be interpreted in two distinct configurations, analogous to the Necker cube, and the vase shadow can induce the classic vase--face perceptual switching. The theoretical models presented in this paper aim to endow machine learning systems with a similar cognitive capability to recognise and interpret such ambiguous patterns in a human-like manner. Building on this, the paper further demonstrates that the so-designed models can be applied to the modelling of social bubbles and other complex phenomena in social networks. The inset illustrates a projective measurement of a qubit using the Bloch sphere.} \label{conceptual}
\end{figure}

Individual perception of ambiguous figures (optical illusions, Fig.~\ref{conceptual}) and collective behaviour in social networks provide a natural experimental setting for this analysis:~they generate rich, repeated decision contexts in which inconsistencies and apparent paradoxes in human choice become both measurable and theoretically informative. My approach uses this setting to articulate a principled connection between quantum formalism, perceptual dynamics and cognitive machine learning.

The term `deterministic decision theories' is used here in its standard sense, referring to models that predict invariant choices under identical conditions. Canonical formulations of expected utility theory (EUT) and cumulative prospect theory (CPT), along with related foundational models, are deterministic in this respect:~a decision-maker who prefers option $A$ to $B$ will, by construction, make the same choice whenever the same decision problem is presented. Empirical evidence from economics and psychology, however, consistently demonstrates choice variability across repeated trials. This discrepancy has motivated the development of stochastic choice models, which introduce probabilistic structure to account for observed behaviour (see, e.g., \cite{fechner1948elements, luce1959individual, blavatskyy2010models}).

Early contributions such as Fechner's model incorporate random perturbations in valuation \cite{fechner1948elements}, while Luce's framework formalises probabilistic choice rules \cite{luce1959individual}. More recent rank-dependent stochastic models extend these ideas to capture systematic response variability.

The distinction between deterministic and stochastic formulations is not merely technical:~it marks a substantive shift in how variability in human decision-making is conceptualised \cite{hey1994investigating, loomes2017preference}. For example, CPT attributes behaviour largely to loss aversion, predicting a tendency to retain the status quo even when switching yields equal or higher expected value. Yet experimental evidence from a relevant study indicates a continuous distribution of responses (with 63\% choosing to stay), rather than a binary outcome~\cite{Mak24_information1}. This suggests that deterministic theories, even when behaviourally enriched, remain incomplete without an explicit stochastic component. Embedding decision-theoretic models within probabilistic frameworks is therefore not optional but necessary \cite{loomes2017preference, Pogrebna_Hills_2026}.

\subsection{Why Novel Approaches are Needed?}
Parallel to these developments, a quantum-theoretic perspective has emerged as a mathematically rigorous alternative for modelling cognition and decision-making \cite{Khr06, Con09, Atm10, Oza20, Bus12, Pot22, Khr_book}. While classical models have struggled to reproduce well-documented anomalies~\cite{LHa10, Aer12, Aer14}, including the Ellsberg paradox \cite{Ell61} and its later refinements \cite{Mac09}, quantum-inspired approaches offer a unified probabilistic language capable of accommodating such effects \cite{bardsley2009experimental}. These models do not replace the traditional decision theory;~rather, they extend it by introducing structure that can capture interference, context dependence and state superposition in cognitive processes~\cite{Bus12}.

The broader intellectual context includes longstanding attempts to relate the computational capacity of the brain to physical principles, including possible quantum-mechanical effects at the neuronal level \cite{Khr06, Koch06, Sur19, Khr20, Khr23, Aer22}. In psychology, analogous efforts have drawn on physical concepts to model perception and cognition \cite{mason2016neural, ludwin2020broken, wong2023seeing}. This has led to the development of quantum models of cognition and decision-making, which have been applied to phenomena such as cognitive dissonance \cite{Fes62}, confirmation bias \cite{Was60} and strategic interaction in games \cite{Pot09, Bus12, Gro17, Pot22}.

The appeal of such models can be clarified by analogy with quantum computation~\cite{Nie02}. A quantum bit (qubit) may occupy the basis states $|0\rangle$ and $|1\rangle$, but also any superposition $\psi = \alpha |0\rangle + \beta |1\rangle$, where $\alpha$ and $\beta$ are complex amplitudes satisfying $|\alpha|^2 + |\beta|^2 = 1$. This formalism allows for a continuum of states, in contrast to the discrete configurations of classical systems. When mapped onto cognition, such representations enable the modelling of intermediate or coexisting mental states, which are difficult to capture within classical probabilistic frameworks \cite{Bus12, Pot22, Mak24_information}.

It is important to distinguish clearly between quantum cognition and the broader class of quantum mind theories. Quantum cognition, which underpins the present work, employs the mathematical structure of quantum mechanics as a modelling framework for probabilistic reasoning and perception~\cite{Bus12, Pot22}. It does not posit that the brain operates as a quantum system. Rather, it uses quantum probability to represent cognitive states, allowing for superposition, interference and context-dependent transitions \cite{Khr06, Bus12, Pot22}. By contrast, quantum mind theories hypothesise that cognitive processes may be physically realised through quantum phenomena in neural substrates, potentially involving entanglement or coherence at the subatomic level \cite{Ham96, Koch06, Kum16, Geo18, Geo20, Geo21, Ker22, Geo22, Geo_book}. While such hypotheses remain speculative, they provide a complementary conceptual backdrop and should not be dismissed outright~\cite{Mak24_APL}.

\subsection{Organisation of the Paper}
The remainder of this paper is organised as follows. First, I examine the origins of the quantum oscillator model within the framework of quantum cognition theory. I then introduce its quantum-tunnelling extension and demonstrate its applicability to modelling bistable perception, using the Necker cube as a canonical example~\cite{Nek32}. 

Subsequently, I show how the quantum oscillator model can be employed to represent social bubbles at the individual level. By incorporating quantum tunnelling, I demonstrate that the extended model captures key features of collective behaviour in social networks. Building on this foundation, I construct quantum-tunnelling networks derived from the oscillator framework. Finally, I illustrate how these networks can be applied to model human perception of optical illusions, analyse sentiment and support machine vision tasks, particularly in the classification of images under conditions of uncertainty.
\begin{figure}
 \centering
 \includegraphics[width=0.79\textwidth]{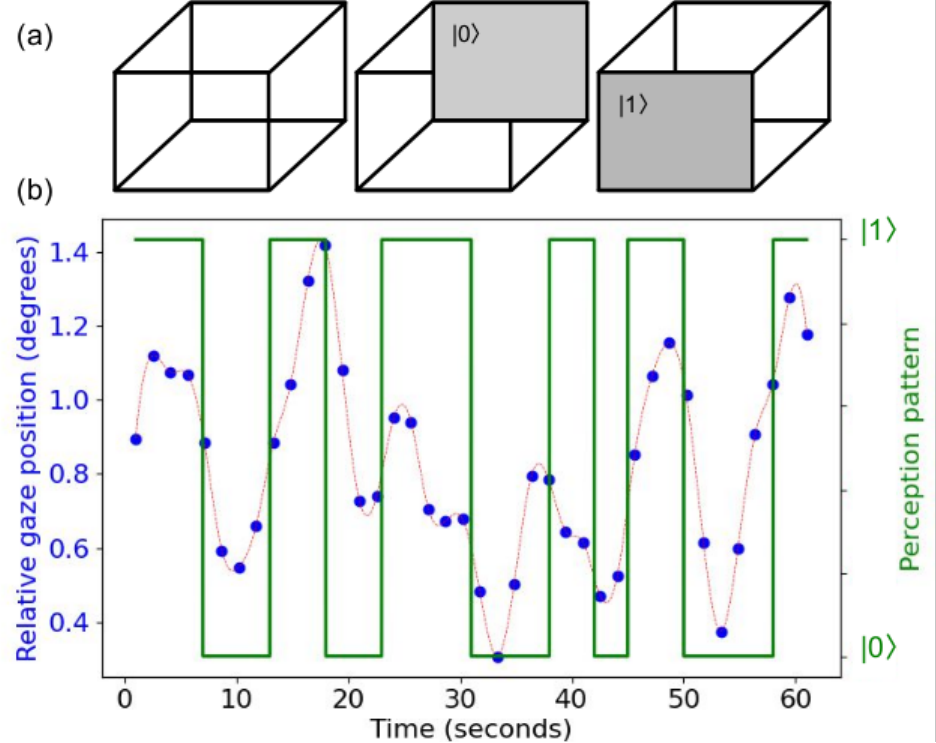}
 \caption{(a)~The Necker cube and its two stable interpretations, denoted as the $|0 \rangle$ and $|1 \rangle$ states in the main text. (b)~Typical experimental discrete perception pattern (green solid line, right $y$-axis), where the cube is perceived in either the $|0 \rangle$ or $|1 \rangle$ state \cite{Cho20}. The continuous dashed line serves as a guide to the eye between the experimental data points (blue dots, left $y$-axis), corresponding to an eye-tracking signal measured simultaneously with the reported perceptual state of the Necker cube \cite{Cho20}.\label{Fig1}}
\end{figure}

\section{Quantum Oscillator Model}
\subsection{Origins of the Model:~the Busemeyer--Bruza Approach}
The original formulation of the quantum-cognitive oscillator model was employed in Ref.~\cite{Bus12} to model bistable perceptual processing of the Necker cube (Fig.~\ref{Fig1}a), a canonical optical illusion arising from an ambiguous graphical representation \cite{Nek32, Lon04} and featured in the Necker video game~\cite{Neckerworld}. In contrast to classical Markov models of bistable perception, where the cube is assumed to occupy one of two discrete states, denoted `0' and `1' \cite{Bus12}, the quantum formulation allows the system to exist in a superposition of the states $|0 \rangle$ and $|1 \rangle$. Upon measurement, the perceptual state undergoes collapse from this superposition to one of the stable interpretations (Fig.~\ref{Fig1}a). 

To illustrate this process, albeit in a somewhat reversed sense compared to the intuitive interpretation, we refer to the computer-generated illustration in Fig.~\ref{conceptual}. While the three-dimensional cube and vase are unique objects, their two-dimensional projections are perceived by an observer as ambiguous figures, namely the Necker cube and the vase--face illusion. Treating these projections as a qubit-like superposition of $|0\rangle$ and $|1\rangle$, we can conceptually project them back into three-dimensional space to recover an unambiguous object corresponding to one of the basis states.

This process is `reverse' because, in the conventional quantum-mechanical picture, a qubit is represented on the Bloch sphere (see the inset in Fig.~\ref{conceptual}) and a measurement projects the state onto one of the coordinate axes, corresponding to the basis states. In that standard interpretation, a superposition collapses to a definite outcome under observation. By contrast, here we begin with an ambiguous, qubit-like superposition inferred from two-dimensional projections and conceptually reconstruct a corresponding three-dimensional, unambiguous state. Hence, rather than modelling the collapse of a known quantum state into a measurement outcome, we infer a definite underlying structure from its ambiguous perceptual manifestation. This approach enables the model to accommodate multiple potential outcomes while operating under a large set of constraints. Its probabilistic outputs have been shown to provide a more efficient description of human mental states than classical models of comparable complexity \cite{Mak24_information1, Mak24_illusions}.

In principle, physical realisations of qubits (for example, using electrons) could be employed to model aspects of human decision-making and perception, with possible implications for the study of complex cognitive processes, particularly if the quantum mind hypothesis were to gain broader empirical support, as discussed in the Introduction section. In practice, however, experimental implementations of quantum systems are technically demanding, especially with respect to measurement~\cite{Kum16, Ker22}. By contrast, quantum systems can be analysed rigorously using established mathematical formalisms, most notably the Schr{\"o}dinger equation~\cite{Kittel, Gri04}. In the present work, I adopt this approach to demonstrate that a mathematically tractable physical model can help bridge the gap between theoretical constructs and experimentally observed psychological phenomena.

The Schr{\"o}dinger equation is a partial differential equation governing the evolution of the wavefunction of a quantum-mechanical system \cite{Kittel, Gri04} (see Appendix~A--D for mathematical details). For a single electron in one spatial dimension, it can be written as
\begin{equation}
  \label{eq:SE}
  \imagi \hbar\frac{\partial \psi(x,t)}{\partial t}=\left[-\frac{\hbar^2}{2m}\frac{\partial^2}{\partial x^2} + V(x)\right]\psi(x, t)\,, 
\end{equation}
where $\psi(x, t)$ is the wavefunction, $\imagi$ is the imaginary unit, $m$ is the electron mass, $\hbar$ is Planck’s constant and $V(x)$ is the potential describing the environment in which the electron evolves.

In Ref.~\cite{Bus12}, separation of variables  applied to Eq.~(\ref{eq:SE}) yields stationary-state solutions that can be interpreted as perceptual probability amplitudes associated with the Necker cube. Within this idealised model, the resulting probabilities exhibit harmonic oscillations. A related perspective is found in Ref.~\cite{Atm10}, where quantum-mechanical principles are used to model the temporal switching between mental states during the perception of ambiguous figures, although in that framework the dynamics is formulated more generally in terms of state evolutions and measurement rather than explicit sinusoidal solutions. (See Appendix~E for an additional discussion of that model.)
\begin{figure}
 \centering
 \includegraphics[width=0.999\textwidth]{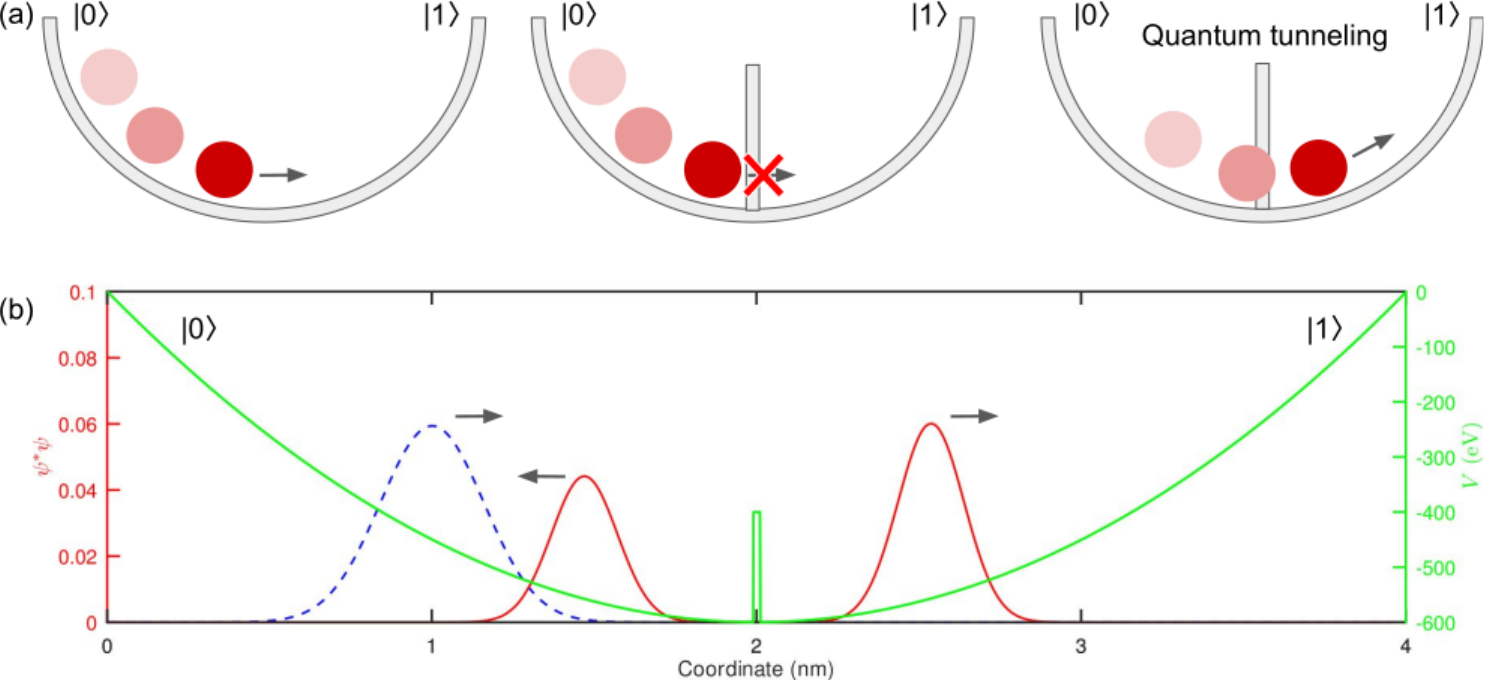}
 \caption{(a)~In classical mechanics, a ball undergoing harmonic motion within a bowl cannot overcome a barrier placed along its path. In contrast, in quantum mechanics an electron confined within a parabolic potential well behaves as a harmonic oscillator and may tunnel through such a barrier. (b)~Numerical simulation of quantum tunnelling through the barrier. The states $|0\rangle$ and $|1\rangle$ correspond to the two perceptual states of the Necker cube.\label{ball_in_the_bowl}}
\end{figure}

\subsection{Quantum Tunnelling Extension of the Oscillator Model:~Necker Cube Example}
As discussed in the Introduction section, a more refined and quantitatively accurate framework, capable of extending beyond the predictive scope of EUT and CPT, can be developed by drawing on quantum-mechanical principles. In particular, quantum-inspired models can be naturally illustrated through the analysis of optical illusions commonly encountered in interactive environments~\cite{wang2021game}.

However, existing quantum-theoretic approaches do not fully account for observations reported in neuroscience studies of the Necker cube and related ambiguous figures. For example, in Refs.~\cite{Run16, Pia17, Joo20}, electroencephalographic signals were recorded alongside subjective reports from observers. While the reported perceptual states exhibit abrupt, seemingly random transitions, the corresponding neural signals display smoothly varying, pulse-like envelopes. Similar continuous dynamics are evident in eye-tracking experiments (Fig.~\ref{Fig1}b; see also Ref.~\cite{Cho20}), where blinks or eye movements are often associated with perceptual reversals \cite{Lon04, Ang20}. Related findings linking perceptual decisions to eye dynamics have been reported in Ref.~\cite{Mat23}. 
\begin{figure}
 \centering
 \includegraphics[width=0.999\textwidth]{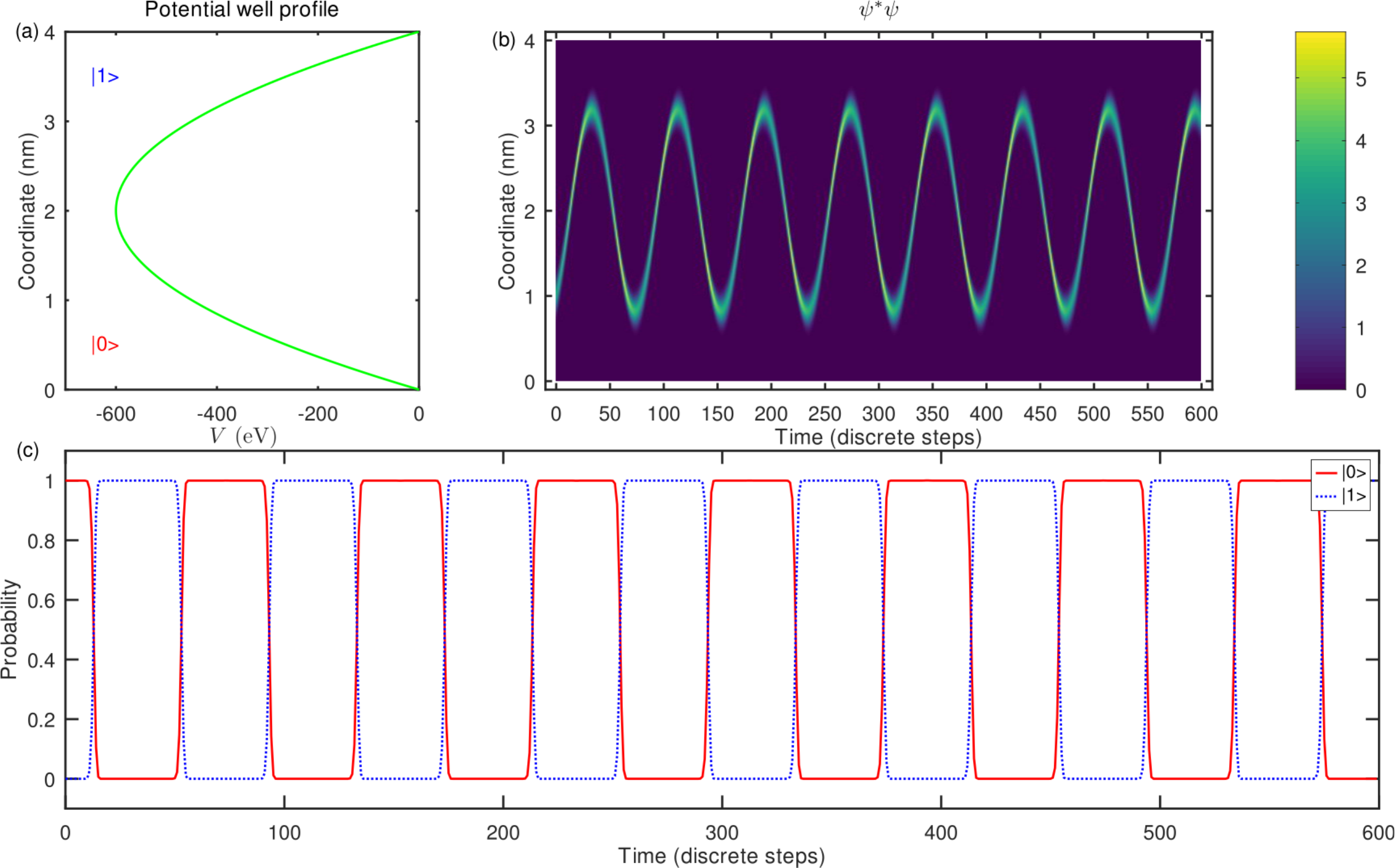}
 \caption{(a)~Model based on a single parabolic potential well. The electron wave packet is initialised at approximately 1\,nm and propagates in the positive direction. (b)~False-colour map of the probability density $\psi^{*}\psi$, shown as a function of spatial coordinate within the well and discrete time. (c)~Probability of finding the electron in the states $|0 \rangle$ (solid curve) and $|1 \rangle$ (dotted curve) as a function of time.\label{parabolic}}
\end{figure}

I emphasise that comparisons with experimental data in the present work are intended as qualitative reference points, as the primary aim is to computationally advance a conceptually novel theoretical framework. A more direct quantitative alignment with experimental observations is feasible~\cite{Mak24_information1}, as also demonstrated by prior studies in which artificial neural networks have successfully reproduced perceptual dynamics associated with optical illusions \cite{Ino94, Ara20, Bat22}.

To model perceptual and behavioural reversals, I consider a physical system in which an electron undergoes harmonic motion in a parabolic potential well (Fig.~\ref{ball_in_the_bowl}a). In classical mechanics, the analogous system is a particle oscillating within a confining potential, such as a ball rolling in a bowl. While a classical particle cannot overcome an barrier without sufficient energy, a quantum particle may traverse such a barrier via tunnelling.

Figure~\ref{ball_in_the_bowl}b presents numerical results for an electron incident on a potential barrier positioned within a parabolic well. The spatial coordinate is shown along the $x$-axis, while the right $y$-axis represents the potential profile $V(x)$. The left $y$-axis shows the probability density, given by the modulus squared of the wavefunction $\psi$, obtained from the solution of Eq.~(\ref{eq:SE}). 

The electron is modelled as a Gaussian wave packet propagating towards the barrier. Upon interaction, part of the wave packet is reflected, while another part is transmitted through the barrier. This behaviour reflects the probabilistic nature of quantum dynamics: there exists a finite probability for both transmission and reflection.

I partition the potential well into two spatial regions, labelled $|0\rangle$ and $|1\rangle$, and interpret these as the two perceptual states of the Necker cube. The probabilities of locating the electron within these regions are taken to represent the probabilities of perceiving the corresponding cube orientations. Initially, the electron is localised in the $|0\rangle$ region, so that $P_{|0\rangle}=1$ and $P_{|1\rangle}=0$, with $P_{|0\rangle}+P_{|1\rangle}=1$. At the conclusion of the simulation, analysis of the reflected and transmitted components yields $P_{|0\rangle}=0.35$ and $P_{|1\rangle}=0.65$.

I first consider a single parabolic potential well (Fig.~\ref{parabolic}a). The numerical results are visualised as a false-colour map (Fig.~\ref{parabolic}b), where colour encodes the probability density as a function of space and time. This representation may be interpreted as a temporal sequence of spatial probability distributions.

The system exhibits harmonic oscillator behaviour, leading to periodic variation in the probabilities of the $|0\rangle$ and $|1\rangle$ states (Fig.~\ref{parabolic}c). Transitions between these states are continuous rather than instantaneous, reflecting temporal nonlocality:~the system evolves smoothly between configurations over finite time intervals \cite{Atm10}. At certain times, the probabilities of the two states become equal, corresponding to a balanced superposition. This behaviour aligns with neurodynamic models of bistable perception, in which perceptual states are temporally interleaved \cite{Noe12}. The results are consistent with previous quantum-theoretic simulations reported in Refs.~\cite{Atm10, Bus12}.

The choice of the initial position of the wave packet warrants comment. Ideally, it should reflect the observer’s preferred initial orientation of the Necker cube, which may vary across individuals and experimental conditions. A fully general treatment would therefore require randomisation of the initial condition, at the cost of increased computational complexity, which is discussed below. For clarity, I fix the initial position at approximately 1\,nm and assume propagation in the positive direction. Alternative initial conditions yield qualitatively similar behaviour and do not affect the main conclusions.
\begin{figure}
 \centering
 \includegraphics[width=0.999\textwidth]{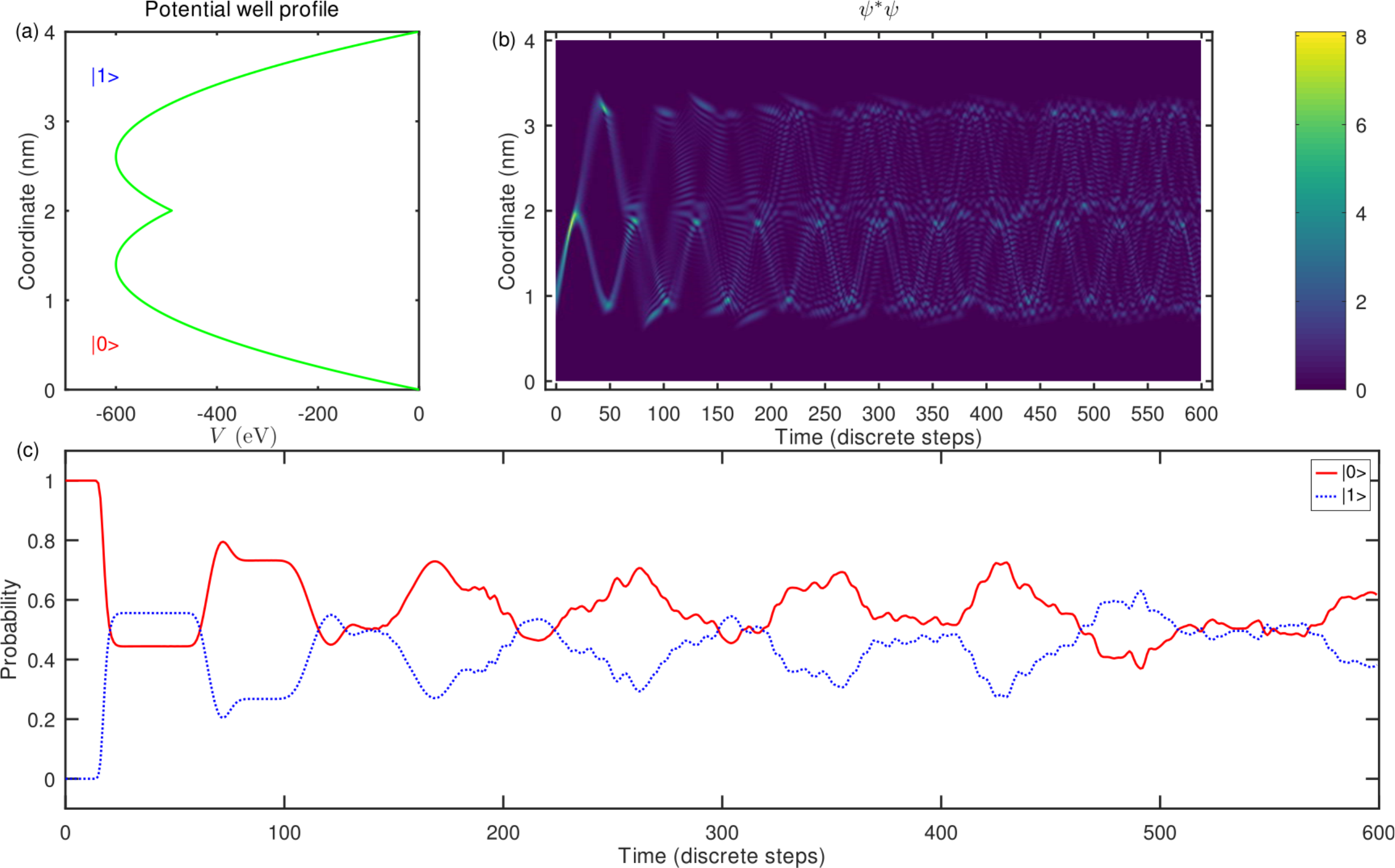}
 \caption{(a)~Model based on a symmetric double parabolic potential well. (b)~False-colour map of the probability density $\psi^{*}\psi$, shown as a function of spatial coordinate within the well and discrete time. (c)~Probability of finding the electron in the states $|0 \rangle$ (solid curve) and $|1 \rangle$ (dotted curve) as a function of time.\label{double_parabolic}}
\end{figure}

Despite capturing key qualitative features, the single-well model does not fully reproduce experimentally observed dynamics (cf.~Fig.~\ref{Fig1}b). To address this limitation, I introduce a double parabolic potential well with a central barrier (Fig.~\ref{double_parabolic}a). In this configuration, quantum tunnelling gives rise to interference effects analogous to those observed in the Young double-slit experiment \cite{Jak64, Suz08, Guo21, Maks25_1}.

This model builds on earlier approaches in which bistable perception was represented using two-level quantum-like systems \cite{Mil09, Con15, Ben18, Atw14}. Related classical-statistical models employing double-well potentials have also been proposed \cite{Pas12, Run16, Mei19, Lep20}, although these do not explicitly solve the Schr{\"o}dinger equation. The potential relevance of quantum tunnelling to bistable perception has been suggested previously \cite{Ben18}, but not developed in a fully dynamical framework.

In many earlier studies, the shapes of the potential wells and barrier heights were chosen phenomenologically. In contrast, our approach emphasises the role of physically grounded parameters. Since tunnelling probability depends sensitively on barrier height, meaningful simulations require careful parameter selection informed by empirical data and modelling considerations.

In Fig.~\ref{double_parabolic}b, the wave packet originates in the left well and propagates towards the barrier. At approximately $T=25$, it splits into two components of comparable amplitude. The resulting state probabilities (Fig.~\ref{double_parabolic}c) exhibit periodic switching between perceptual states. Notably, except for an initial transient regime, the system predominantly occupies superposition states due to interference effects. Such behaviour is consistent with the notion of cognitive superposition discussed in Refs.~\cite{Atm10, Bus12}.

These results are also consistent with experimental findings indicating that perceptual states may become unstable prior to a reported reversal \cite{Wil23}. While observers can exert partial control over perceptual switching (for example, by modulating attention) the reversals cannot be entirely suppressed \cite{Lon04}. The model reproduces both the variability in reversal frequency and the irreducibility of the reversal process.

\begin{figure}
 \centering
 \includegraphics[width=0.999\textwidth]{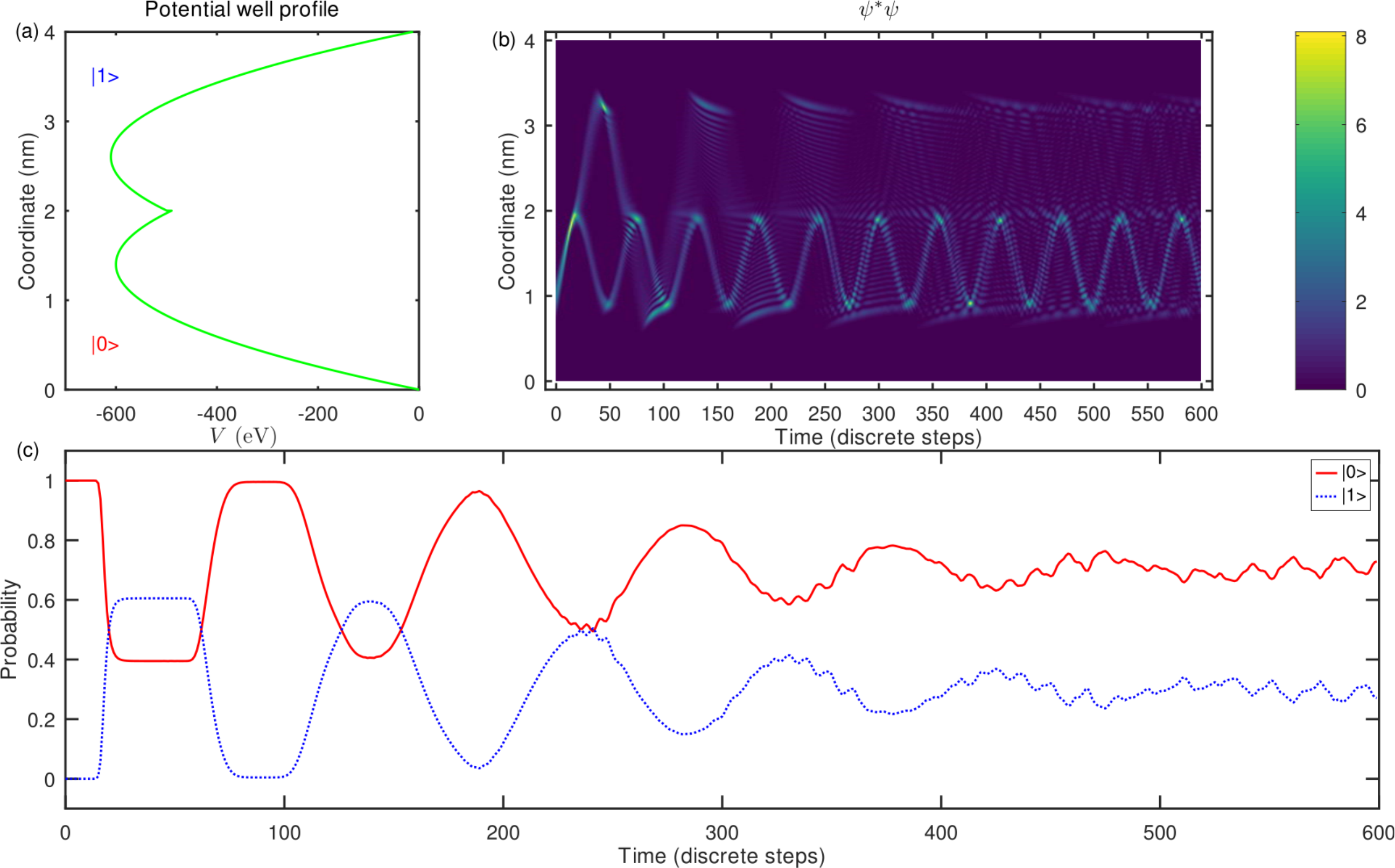}
 \caption{(a)~Model based on an asymmetric double parabolic potential well, obtained by shifting one of the wells by $-10$\,eV. (b)~False-colour map of the probability density $\psi^{*}\psi$, shown as a function of spatial coordinate within the well and discrete time. (c)~Probability of finding the electron in the states $|0 \rangle$ (solid curve) and $|1 \rangle$ (dotted curve) as a function of time.\label{asym_double_parabolic}}
\end{figure}

Perceptual ambiguity is typically asymmetric: observers tend to favour one interpretation over another \cite{Nak11}, and reversal rates depend on familiarity, adaptation and attentional factors \cite{Lon04, Sto12}. These effects can be incorporated by modifying the potential landscape. For example, an asymmetric double well with an increased barrier height (Fig.~\ref{asym_double_parabolic}a) reduces tunnelling probability, thereby favouring localisation in the initial well (Fig.~\ref{asym_double_parabolic}b,~c). This behaviour may be interpreted as a bias induced by gaze control or reduced blinking \cite{Lon04}. Nevertheless, the model confirms that complete suppression of perceptual reversals is not achievable, as the system inevitably transitions between states over time.

\section{Quantum Oscillator in Social Network Modelling}
In this section, I demonstrate how the potential-well quantum oscillator model can be extended to capture social behaviour. I begin with a simple and fundamental case of a single potential well, here taken as rectangular, which preserves the underlying physics while illustrating that alternative well configurations are both mathematically and computationally tractable within quantum cognition theory.

I then increase the number of wells to transition from modelling individual behaviour to collective dynamics, representing a social network as a sequence of wells and barriers. Alternatively, this structure may be interpreted as a single potential well with multiple internal barriers. Within this model, I show that the quantum oscillator model can be applied to address a range of problems for which quantum cognition theory offers meaningful explanatory power. 

\subsection{Modelling of Social Bubbles}
In an age shaped by social media and algorithmic curation, individuals are increasingly exposed to information streams that reinforce existing beliefs while attenuating opposing viewpoints~\cite{Bai18, Gal22}. This phenomenon, variously described as a social bubble or echo chamber, has significant consequences for the formation, persistence and polarisation of both individual and collective opinions~\cite{Cin21, Gal24, Gal_book}. Within such closed informational environments, consensus is often driven less by evidence or deliberation than by repetition and social reinforcement. Over time, this process can produce systematic bias, ideological drift and resistance to correction, even in the presence of compelling contrary evidence~\cite{Bai18, Cin21, Gal_book, Khr23}.
\begin{figure}
 \centering
 \includegraphics[width=0.999\textwidth]{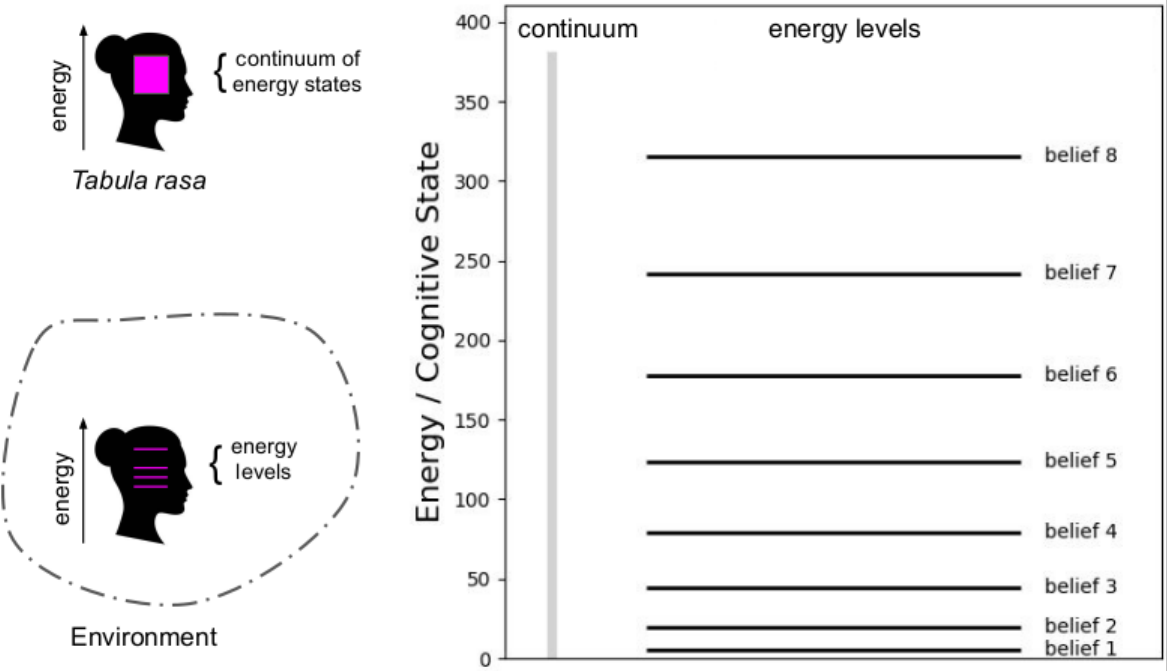}
 \caption{Illustration of the {\it tabula rasa} model, where an initially continuous spectrum of cognitive (energy) states becomes structured by the social environment, yielding discrete belief states and a corresponding propensity for ideology, bias and opinion polarisation.\label{social_bubble_MDPI}}
\end{figure}

Classical models of opinion dynamics, typically grounded in rational choice theory and linear diffusion processes, are generally ill-equipped to capture these effects~\cite{Gal_book}. They assume that individuals update beliefs in a consistent and logically coherent manner, akin to Bayesian agents~\cite{Szn00, Red19}. Empirical evidence, however, suggests a more complex picture:~human reasoning is strongly mediated by affect, identity and social context~\cite{Pot22, Khr23}. Individuals may discount or reject accurate information when it conflicts with prior beliefs or threatens group affiliation~\cite{Bai18}.

The quantum oscillator model provides a mathematically coherent alternative for modelling such behaviour. By representing belief states as superpositions---coexisting and potentially incompatible, probabilistic states---it holds the potential to accommodate ambiguity, contextual dependence and cognitive tension~\cite{Pogrebna_Hills_2026}. A decision or expressed opinion can then be understood as the outcome of a context-dependent `measurement' that selects one of several latent possibilities. This formalism naturally captures the fluid and often non-committal nature of human belief prior to explicit judgement~\cite{Tac20, Mak24_gender}.

In social settings, the act of communication itself functions as a measurement process~\cite{Bai18}. Interactions, whether conversations, social media exchanges or exposure to news, serve as contextual operators that shape the manifestation of latent beliefs. Within a tightly coupled informational environment, repeated interactions of this kind progressively constrain the accessible belief space. The result is a narrowing of cognitive diversity and the emergence of correlated belief structures across individuals, even in the absence of explicit coordination.

Quantum models of opinion dynamics thus offer a principled account of how consensus, conformity and polarisation arise from the interplay between individual cognition and social influence~\cite{Khr23}. They also point towards mitigation strategies, including the deliberate introduction of informational diversity, structured cross-group engagement and the design of recommendation systems that preserve, rather than suppress, cognitive heterogeneity~\cite{Pot22}.

To illustrate these ideas, consider a simple and idealised model (Fig.~\ref{social_bubble_MDPI}, top left) that I term \textit{tabula rasa}. In philosophical usage, this denotes the notion of a `blank slate', whereby an individual begins without predefined mental content and acquires knowledge through experience. I adopt this concept as a physical analogy. An individual is modelled as a quantum-like entity described by a wavefunction, while information is associated with the energy states of the system~\cite{Toy10, Dit14, Vop19}. In the absence of environmental constraints, the system admits a continuum of states, corresponding to an unconstrained cognitive landscape.

The introduction of an environment imposes structure on this space (Fig.~\ref{social_bubble_MDPI}, bottom left). Mathematically, this can be represented by confining the system within a bounded domain, such as a one-dimensional infinite potential well of width $L$~\cite{Kittel}. The resulting boundary conditions require the wavefunction to vanish at the domain boundaries, permitting only standing-wave solutions. Consequently, the allowed states become discrete rather than continuous, mirroring the formation of stable belief structures through experience and social conditioning.

The permitted wavevectors are given by
\begin{equation}
k_n = \frac{n \pi}{L}, \qquad n = 1, 2, 3, \dots,
\end{equation}
with corresponding stationary states
\begin{equation}
\psi_n(x) = \sqrt{\frac{2}{L}} \, \sin\!\left(\frac{n \pi x}{L}\right), \qquad 0 < x < L.
\end{equation}
The associated energy levels are
\begin{equation}
E_n = \frac{\hbar^2 k_n^2}{2 m} = \frac{n^2 \pi^2 \hbar^2}{2 m L^2} = \frac{n^2 h^2}{8 m L^2}, \qquad n = 1, 2, \dots,
\end{equation}
where $m$ is the effective mass, $L$ the system size and $\hbar = h/2\pi$ the reduced Planck's constant.

From a physical perspective, confinement restricts the admissible wavelengths to those compatible with the imposed boundary conditions, yielding a discrete spectrum (Fig.~\ref{social_bubble_MDPI}, right). As $L$ increases, the spacing between energy levels decreases as $\Delta E \propto 1/L^2$, and the spectrum approaches a continuum in the limit $L \to \infty$. Conceptually, this corresponds to a loosening of environmental constraints and a recovery of cognitive openness.
\begin{figure}
 \centering
 \includegraphics[width=0.7\textwidth]{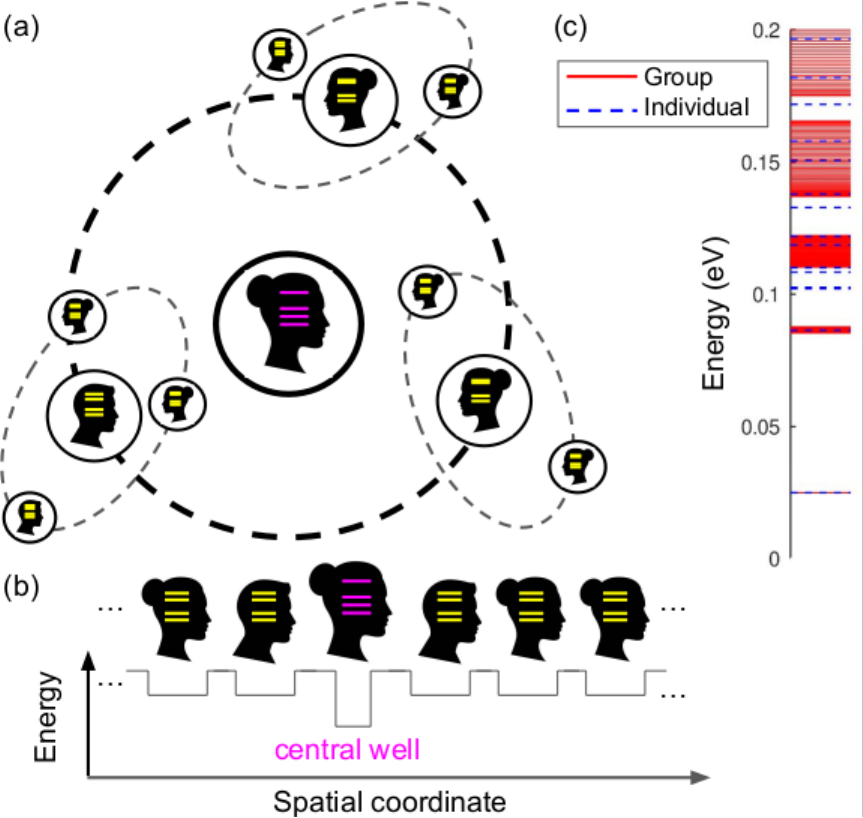}
 \caption{(a)~A physics-inspired model of a social network, in which individuals---represented as quantum particles in potential wells---interact (i.e.~socially and physically), leading to modifications of their energy-level structures (i.e.~beliefs and opinions) in accordance with quantum-mechanical principles. (b)~A one-dimensional, computationally tractable realisation of the network shown in panel~(a). (c)~Numerically calculated evolution of energy levels from an individual to a collective belief structure.\label{social_network_MDPI}}
\end{figure}

More generally, alternative confining potentials may be introduced to represent different forms of social or informational structure. For instance, a parabolic potential discussed in the previous sections yields equally spaced energy levels, corresponding to uniformly structured cognitive states. Here, unlike the infinite well, confinement arises from a restoring force, and the wavefunction is localised by the potential itself rather than by hard boundaries. This leads to a discrete set of allowed states given by~\cite{Kittel}
\begin{equation}
\psi_n(x) = \frac{1}{\sqrt{2^n n!}} \left(\frac{m\omega}{\pi \hbar}\right)^{1/4}
H_n\!\left(\sqrt{\frac{m\omega}{\hbar}}\,x\right)
\exp\!\left(-\frac{m\omega x^2}{2\hbar}\right), \qquad n = 0,1,2,\dots,
\end{equation}
where $H_n$ are the Hermite polynomials and $\omega$ is the angular frequency of the oscillator. The corresponding energy levels are quantised as
\begin{equation}
E_n = \hbar \omega \left(n + \frac{1}{2}\right), \qquad n = 0,1,2,\dots,
\end{equation}
where $m$ is the effective mass and $\hbar$ is the reduced Planck's constant. 

\subsection{Models of Opinion Formation in Social Networks}
Thus, as demonstrated above, in contrast to the infinite potential well, the energy spectrum is therefore evenly spaced, with constant level separation $\Delta E = \hbar\omega$, reflecting the harmonic nature of the confining potential and certain advantages for quantum-cognitive modelling. However, a practical advantage of the rectangular potential is its relatively straightforward mathematical and computational tractability:~to a large extent, it relies on formal tools that are already well established in fields such as quantitative finance and behavioural science. Importantly, using the same methods and just slightly increasing the complexity of the underlying computations, one can combine several identical or similar well into a network that models human behaviour in social media.

This idea is illustrated in Fig.~\ref{social_network_MDPI} and validated through rigorous numerical simulations using a finite-difference method outlined in Appendix~B. For clarity, I initially arrange individuals within an orbital-like representation of a social network (Fig.~\ref{social_network_MDPI}a). While this construction admits a formal physical interpretation, its primary purpose is illustrative:~to show that, analogous to quantum particles, individuals interact and the discrete energy levels associated with them in a social bubble are modified through mutual interactions.

However, such a model is mathematically challenging, as it is generally three-dimensional, involves many degrees of freedom and may include potential wells of varying geometry. To retain analytical and computational tractability, I restrict the formulation to one dimension using rectangular wells (Fig.~\ref{social_network_MDPI}b), while preserving the underlying quantum-cognitive approach. This reduced model remains readily solvable and the results of the corresponding computations are presented in Fig.~\ref{social_network_MDPI}c.

As can be seen, the initially discrete energy levels (beliefs) of an isolated individual within their own social bubble (blue dashed lines in Fig.~\ref{social_network_MDPI}c) become modified when that individual is embedded within a broader social network (red solid lines in Fig.~\ref{social_network_MDPI}c). This process is inherently reciprocal and complex, governed by quantum-mechanical principles~\cite{Mak24_information}. As a result, the spectrum evolves into a quasi-continuous band structure, interspersed with non-occupied (physically forbidden) energy ranges, such as those observed around 0.1\,eV in Fig.~\ref{social_network_MDPI}c.

It has been argued in Ref.~\cite{Mak24_information} that the emergence of forbidden gaps can be used to model opinion polarisation and related social phenomena, whereby individual positions become entrenched and resistant to change, often leading to counterproductive outcomes such as the backfire effect~\cite{Bai18}. The rationale behind this proposal is that modifying these forbidden energy bands requires additional energy in the physical sense, with energy being equivalent to information~\cite{Toy10, Dit14, Vop19}.

This hypothesis can be tested by performing calculations with otherwise identical potential wells, except for a single perturbed well, thereby introducing a conflicting piece of information into an otherwise coherent and aligned group of individuals. This approach yields results that are at least qualitatively consistent with patterns observed in domains such as politics~\cite{Bai18}, vaccination attitudes~\cite{nyhan2015does}, climate change perception~\cite{dixon2019unintended} and abortion debates~\cite{liebertz2021backfiring}, highlighting a range of societal challenges that may be amenable to modelling within this model~\cite{Mak24_information}.

\subsection{Comparison with Khrennikov's Social Laser Model}
As demonstrated throughout this paper, quantum-cognitive models based on quantum tunnelling and related quantum-mechanical effects, implemented via numerical solutions of the Schr{\"o}dinger equation, are broadly consistent with, and in many cases formally equivalent to, established approaches in the field~\cite{Khr06, Atm10, Bus12, Aer14, Pot22}. These connections extend naturally to adjacent frameworks, including models of open quantum systems and the notion of social laser dynamics~\cite{Khr23}, as well as to classical descriptions of social behaviour~\cite{Gal_book}.

Khrennikov's social laser model~\cite{Khr16_laser} adopts a quantum-like formalism at the collective level, where coherence, amplification and synchronisation of information states across interacting agents give rise to macroscopic social phenomena. Thus, a coherent interpretation emerges by viewing the discrete energy states model and social laser approach as complementary descriptions operating at different scales. In fact, the discrete energy-level model captures the `microdynamics' of individual cognition, where transitions between localised states encode perception, judgement and learning. The social laser framework, by contrast, describes `mesoscopic' and `macroscopic' regimes in which interactions between individuals lead to collective coherence and amplification effects. Therefore, individual cognitive states may be regarded as the elementary units whose interactions, when sufficiently coupled, give rise to the collective phenomena described by the social-laser picture.

Such a multiscale perspective suggests a natural integration pathway of the two models:~the energy landscape governing individual cognition defines the local structure of belief states, while interactions between agents reshape this landscape and enable coherence across the network. This unified view preserves the interpretability of discrete quantum-cognitive states while extending their applicability to collective behaviour, thereby bridging individual decision-making and large-scale social dynamics within a common formal model.

\section{Quantum-Cognitive Neural Networks}
In this section, building on the network of potential barriers introduced previously, through which an electron can tunnel, thereby modelling human behaviour, I demonstrate that the incorporation of deep neural network architectures leads to a quantum–cognitive neural network grounded in quantum cognition theory. Such an approach, subject to further experimental validation, has the potential to serve as an AI model that more closely mimics human cognition.

I show that this type of network can be applied to machine vision tasks, particularly in recognising optical illusions in a manner consistent with human perception, an area that remains challenging for conventional neural networks~\cite{Sha24, Zha24, Zha24_1}. I also demonstrate that the model can be used for standard object recognition tasks under conditions of uncertainty, where classical machine learning methods often struggle.
\begin{figure}
 \centering
 \includegraphics[width=0.999\textwidth]{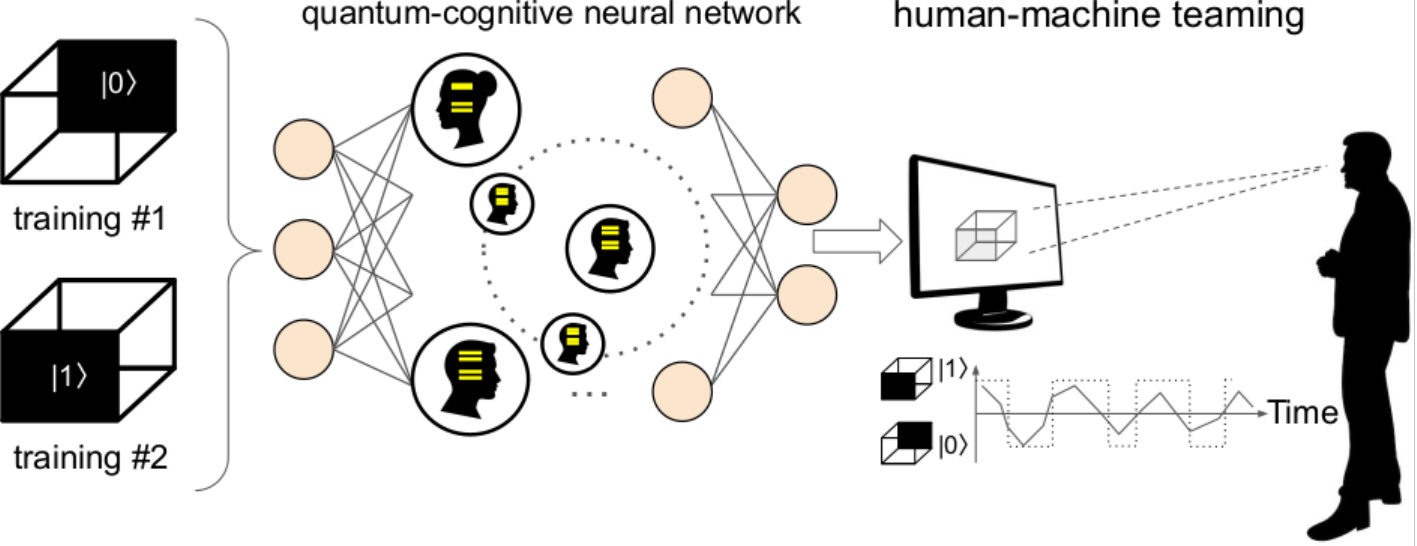}
 \caption{Conceptual illustration of a quantum-cognitive neural network employing quantum tunnelling effects and the principles of quantum cognition introduced in the preceding sections. In this example, the network is trained on the two fundamental perceptual states of the Necker cube. When presented with an ambiguous Necker cube stimulus, the network’s predictions can be recorded as a function of time (cf.~the bottom-right inset of the present figure with Fig.~\ref{Fig1}b) and compared with human responses, where participants report their perceived interpretation of the ambiguous figure. The proposed network can be trained on other ambiguous figures, and more generally on images of real-world objects under conditions of uncertainty and ambiguity.\label{neural_network_MDPI}}
\end{figure}

\subsection{Algorithm}
\subsubsection{General Discussion}
I begin with a conceptual overview, followed by a formal description of the algorithm and the specific neural network employed in this work.

The conceptual framework is illustrated in Fig.~\ref{neural_network_MDPI}, where the network is exemplified using the Necker cube optical illusion. The network, depicted schematically by interconnected nodes linked to stylised representations of human cognition via discrete energy levels, embodies quantum oscillators with tunnelling. Structurally, it retains the standard architecture of deep neural networks~\cite{Kim17}, comprising input, hidden and output layers.

Activation functions in neural networks are mathematical mappings that determine whether a neuron is activated, thereby introducing nonlinearity into the model~\cite{Kim17}. This nonlinearity is essential, as it enables the network to capture complex relationships and patterns in data, transforming an otherwise linear system into a powerful system capable of modelling high-dimensional inputs such as images and text.

In the model shown in Fig.~\ref{neural_network_MDPI}, learning is mediated through quantum tunnelling. Specifically, the probability of an electron tunnelling through a potential barrier governs the activation of a neuron. Although the overall formulation remains mathematical and the underlying quantum process is linear, the transmission characteristics of tunnelling are intrinsically nonlinear (see Appendix~C). It is therefore plausible to hypothesise that such a mechanism can emulate aspects of human cognition, consistent with the quantum-cognitive and oscillator-based models introduced above.

In the particular realisation shown in Fig.~\ref{neural_network_MDPI}, the two fundamental perceptual states of the Necker cube are disambiguated by rendering one of the cube faces opaque. (Other forms of disambiguation, including variations in line thickness and colour coding, were also tested and produced consistent results, albeit at a higher computational cost.) For convenience, these two states are labelled as $|0\rangle$ and $|1\rangle$, respectively. Note that these labels do not form part of the graphical field encoding the pixel data of the images; rather, they serve as class labels used during training, whereby the network is presented with batches of images corresponding to state $|0\rangle$ or $|1\rangle$.

As in standard image recognition datasets such as MNIST and Fashion-MNIST~\cite{Den12, Xia17}, the pixel values of the cube images are converted into a numerical representation and appropriately normalised following established machine-learning practice~\cite{Kim17}. The network is then trained in the usual manner, treating the images analogously to handwritten digits or other standard inputs. This feature is important, as it allows the same model to be readily extended to real-world applications.

Once trained, the network is evaluated by repeatedly presenting it with the same ambiguous Necker cube image, encoded in the same format as the disambiguated training data. The output is a sequence of predicted states taking values between 0 and 1, corresponding to the fundamental perceptual states $|0\rangle$ and $|1\rangle$. These predictions form a time-dependent signal that oscillates between the two states (see the illustration in the bottom-right inset in Fig.~\ref{neural_network_MDPI}).

At this stage, the neural network can be integrated into a human--machine teaming system, in which a human observer simultaneously views the Necker cube and reports perceptual changes, for example by pressing a button, or is monitored via eye-tracking or neural activity measurements~\cite{Joo20, Cho20}. Such a system has the potential to advance research in psychology and decision-making, while also providing valuable data for neuroscientists. 

\subsubsection{Software Implementation}
The neural network architecture implemented in software comprises an input layer with $L=100$ nodes, three hidden layers each containing $N=20$ nodes and an output layer with $M=2$ nodes used to classify the perceptual state of the Necker cube. The network weights are updated using a cross-entropy-based back-propagation algorithm~\cite{Kim17}, with a learning rate parameter $\alpha$. As discussed below, these weights can be initialised using standard techniques based on pseudo-random number generators or, alternatively, by employing quantum-generated random numbers.

For clarity, I first describe the conventional version of this neural model and then demonstrate how it can be extended to incorporate quantum tunnelling. Both versions share an otherwise identical topology.

In the traditional version, the activation function of the hidden-layer nodes is the Rectified Linear Unit (ReLU)~\cite{Kim17}
\begin{equation}
  \phi_{ReLU}(x_j) = 
  \begin{cases}
      x, & \text{$x_j>0$}\\
      0, & \text{$x_j\leq0$}
    \end{cases}\,,
  \label{eq:ReLU}
\end{equation}
where $j=1 \dots L$ denotes the input node index and $x_j$ is the corresponding input. The output layer employs the Softmax function~\cite{Kim17}
\begin{equation}
  \phi_{smax}(v_i) = \frac{\exp(v_i)}{\sum_{k=1}^{M} \exp(v_k)}\,,
  \label{eq:softmax}
\end{equation}
where $v_i$ is the weighted input to the $i$th output node and $M$ is the number of output nodes. This ensures the normalisation condition $\sum_{k=1}^{M}\phi_{smax}(v_k)=1$.

The network is trained as follows~\cite{Kim17}:
\begin{enumerate}
\item Define two output nodes corresponding to the perceptual basis states of the Necker cube, $|0\rangle = [1~0]$ and $|1\rangle = [0~1]$;
\item Initialise all network weights randomly in the interval $[-1,1]$;
\item Provide the input data $x_j$ together with the corresponding target outputs $d_i$;
\item Compute the output error
\[
e_i = d_i - y_i,
\]
where $y_i$ is the network output;
\item Back-propagate the error by setting $\delta_i = e_i$ at the output layer and computing
\[
e_i^{(n)} = W^{(n)\top}\delta_i, \qquad
\delta_i^{(n)} = \phi_{\mathrm{ReLU}}^{\prime}\!\left(v_i^{(n)}\right) e_i^{(n)};
\]
\item Propagate the error through all hidden layers;
\item Update the weights according to
\[
w_{ij}^{(n)} \leftarrow w_{ij}^{(n)} + \Delta w_{ij}^{(n)}, \qquad
\Delta w_{ij}^{(n)} = \alpha\, \delta_i^{(n)} x_j;
\]
\item Repeat for all training samples;
\item Iterate until convergence.
\end{enumerate}

Unlike the traditional network, the activation function $\phi_{QT}$ of the hidden-layer nodes of the quantum-tunnelling version is defined by the algebraic expressions for the transmission coefficient $T$ of an electron tunnelling through a potential barrier. These expressions are well established~\cite{McQ97, Geo18} and are summarised in Appendix~C for self-consistency. The output nodes of the network continue to employ the Softmax function, Eq.~(\ref{eq:softmax}).

Training and inference proceed as follows. First, the output nodes are defined to represent the target labels of the training dataset. The network weights are initialised in the range $[-1,1]$ using a random number generator. Given input data $x_j$ and corresponding targets $d_i$, the error is computed as $e_i=d_i-y_i$, where $y_i$ is the network output. The error signal $\delta_i=e_i$ is then back-propagated through the network to compute the hidden-layer parameters $\delta_i^{(n)}$ using
\[
e_i^{(n)} = W^{{(n)}^\top}\delta_i,\qquad
\delta_i^{(n)} = \phi_{QT}'\!\left(v_i^{(n)}\right)e_i^{(n)},
\]
where $n$ denotes the hidden-layer index, the prime indicates differentiation with respect to the argument and $W^\top$ is the transpose of the corresponding weight matrix.

In both models, the inference stage mirrors the forward-pass steps of the training procedure~\cite{Kim17}. In practice, 1000 epochs are sufficient to achieve convergence.
\begin{figure}[t]
\centering
 \includegraphics[width=0.8\textwidth]{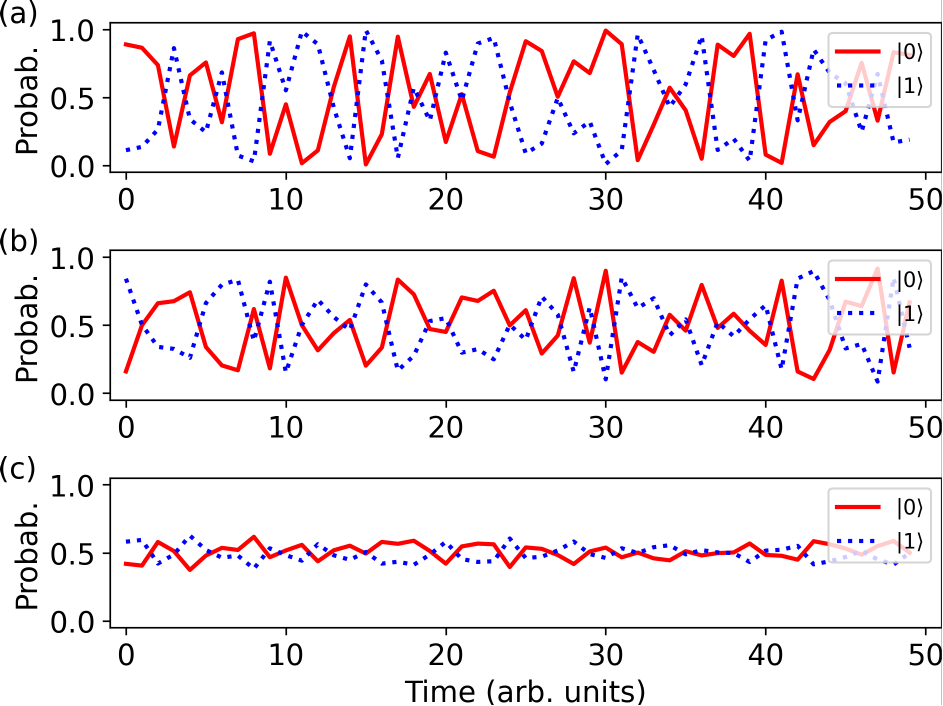}
 \caption{Perceptual switching curves produced by the quantum-tunnelling neural network trained on the Necker cube. The data points with probability $P_{|0\rangle}=0$ or $P_{|1\rangle}=1$ correspond to the fundamental perceptual states of the Necker cube. The remaining data points are in a superposition of states $|0\rangle$ and $|1\rangle$ with $P_{|0\rangle}+P_{|1\rangle}=1$.\label{Fig_Necker_result}}
\end{figure}

\subsection{Applications in Machine Vision}
In what follows, I present the results of applying the quantum-cognitive neural network to the Necker cube (and, Appendix~D, to Rubin’s vase, another paradigmatic optical illusion widely used in psychology and quantum cognition studies~\cite{Kha21_1}). I then evaluate the model on a real-world ambiguous machine vision dataset. 

\subsubsection{Optical Illusions}
Figure~\ref{Fig_Necker_result}a reveals time-dependent switching between the fundamental perceptual states $|0\rangle$ and $|1\rangle$. This switching is not abrupt, but proceeds via intermediate states corresponding to superpositions of $|0\rangle$ and $|1\rangle$. A similar pattern is evident in Fig.~\ref{Fig_Necker_result}b for a thicker potential barrier. Although the peak probabilities of the two percepts decrease to approximately 0.85, the neural network remains capable of reliably distinguishing between the two states of the Necker cube.

A further increase in the barrier thickness (Fig.~\ref{Fig_Necker_result}c) reduces the contrast between the probabilities associated with $|0\rangle$ and $|1\rangle$. In this regime, the neural network exhibits behaviour analogous to human perception \cite{Kor12}: while ambiguous visual information can be processed rapidly through a few recurrent neural cycles \cite{Lam00}, additional time is required to stabilise and report a definite percept.

Indeed, I find that increasing the number of training epochs to 2000 yields a markedly more separable perceptual pattern, indicating that the model effectively requires more time to reach a decision. This behaviour is particularly relevant for modelling perceptual differences across demographic groups \cite{Sun97, lo2011investigation}. For example, since the dynamics of eye blinking, and hence visual information processing, tend to slow with age \cite{Sun97, Yan24}, the neural network with thicker potential barriers provide a plausible proxy for modelling perception in older or visually impaired observers.

It is worth noting that the oscillatory behaviour produced by the neural network in Fig.~\ref{Fig_Necker_result} qualitatively agrees with the predictions of the quantum-tunnelling oscillator model shown in Fig.~\ref{parabolic}, Fig.~\ref{double_parabolic} and Fig.~\ref{asym_double_parabolic}. In particular, the neural-network results combine the periodic behaviour observed in Fig.~\ref{parabolic} with the oscillating superposition of perceptual states seen in Fig.~\ref{double_parabolic} and Fig.~\ref{asym_double_parabolic}.

This correspondence can be better understood by recalling that purely periodic oscillations in the oscillator model arise in the absence of a barrier, as in the original Busemeyer--Bruza formulation discussed at the beginning of this paper. By contrast, introducing a barrier and allowing for tunnelling, together with its dependence on barrier thickness, effectively combines oscillatory dynamics with decay, driving the system towards a partially indeterminate state. This behaviour is precisely reflected in Fig.~\ref{Fig_Necker_result}b,~c, now realised within the neural-network framework. Thus, the conceptual chain linking the Busemeyer--Bruza model of quantum cognition to the quantum-tunnelling oscillator model, and further to quantum-tunnelling neural networks, indicates that the resulting neural architecture retains the essential formalism of quantum cognition. It can therefore be legitimately regarded as an extension of quantum cognition theory.
\begin{figure}
 \centering
 \includegraphics[width=0.999\textwidth]{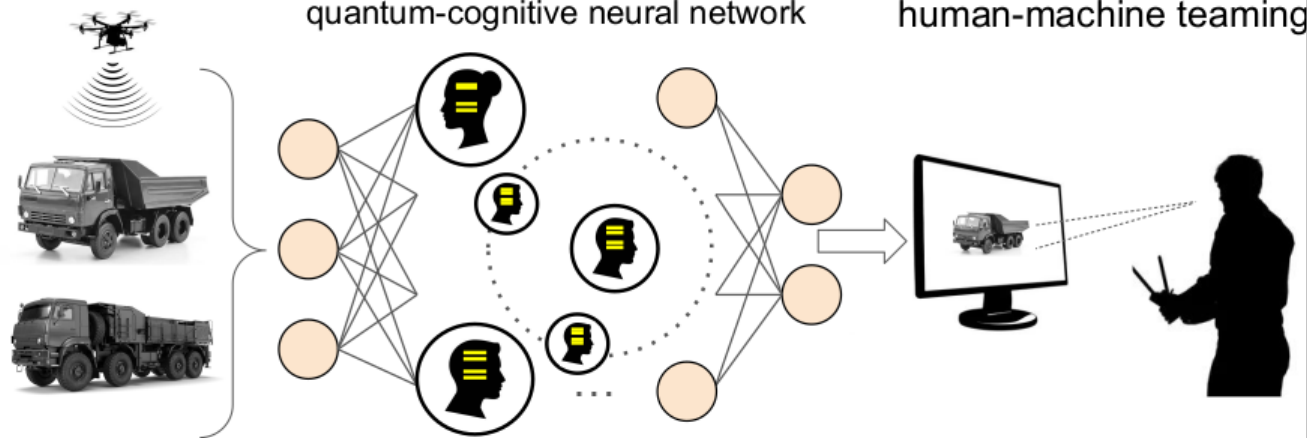}
 \caption{Conceptual redrawing of Fig.~\ref{neural_network_MDPI} using visually ambiguous images of civilian and military trucks operating in a combat zone. This particular illustration employs computer-generated images of civilian and military Kamaz trucks that share highly similar cabin and chassis designs. Other makes of European-produced military trucks deployed in conflict zones exhibit similar design commonality with their civilian counterparts. When captured by drones under conditions such as fog, smoke, rain and related visual distortions, these vehicles may be misclassified even by a highly trained, and potentially psychologically biased, human operator. By treating such ambiguous vehicle images analogously to optical illusions and integrating human--machine teaming, the neural network can assist in more reliable classification, thereby contributing to life-saving decision-making.\label{trucks_neural_network_MDPI}}
\end{figure}

\subsubsection{Classification of Real-World Ambiguous Objects}
Previous relevant work~\cite{Pogrebna_Hills_2026, Maks25, Maks25_1} has demonstrated a strong physical connection between the quantum tunnelling effects central to this study and the fundamental phenomenon of the double-slit experiment~\cite{Nie02, Gri04}. A psychological analogue of the double-slit experiment was proposed in Ref.~\cite{Bus12} and further developed in Ref.~\cite{Wan16}, where a quantum-cognitive framework was used to model decision-making tasks such as categorising human faces as `good' guy or `bad' guy and deciding to act `friendly' or `aggressive'.

Given that the quantum-tunnelling neural network developed here retains the functional structure of quantum-cognitive models, it can, in principle, be applied to analogous classification problems. Motivated in part by the author's background and the ongoing war in his country of origin, one test case considered in this work involves classifying transport vehicles as military or civilian under conditions of uncertainty and ambiguity, as commonly encountered in conflict zones.

The primary rationale of this exercise is not the advancement of military technologies but measures to protect civilians. The use of drone warfare has transformed modern military practice, including the emergence of so-called `kill zones' that are continuously patrolled by drones and in which any moving vehicle may be treated as a target. Given the presence of civilians and the use of civilian vehicles for military purposes, human-operated drone systems have resulted in significant casualties among non-combatant populations.

A customised machine-vision dataset of military and civilian vehicles used in conflict zones was created based on open-source imagery~\cite{Maks25_2} including the well-established CIFAR library~\cite{CIFAR}, and the neural network was trained on this dataset. Subsequently, following a procedure analogous to that used for the Necker cube, the classification performance of the network was evaluated. In this assessment, particular attention was paid to misclassification events, as these directly determine whether a drone might incorrectly strike a civilian vehicle instead of a military target.
\begin{figure}
 \centering
 \includegraphics[width=0.999\textwidth]{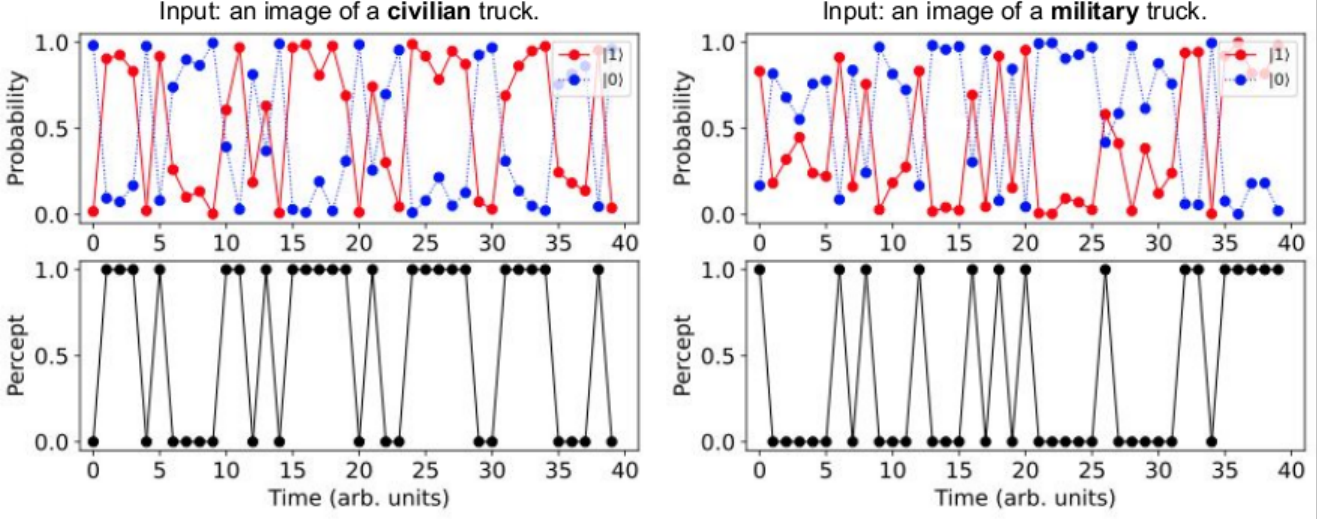}
 \caption{Performance of the quantum-tunnelling deep neural network on images of civilian and military trucks, presented in a format analogous to that used for the Necker cube and Rubin's vase optical illusions. The top panels show the quantum probabilities associated with the model outputs, while the bottom panels depict the corresponding classical percept obtained via thresholding, where $\mathrm{Percept} = 1$ if $P_{\ket{1}} \geq 0.5$ and $\mathrm{Percept} = 0$ otherwise. For a civilian truck input (left), the model yields a higher probability of the `civilian' class, reflected by a predominance of $\mathrm{Percept}=1$ outcomes. Conversely, for a military truck input (right), the model favours the `military' class, with the classical percept predominantly taking the value $\mathrm{Percept}=0$.}
\label{confused_trucks_MDPI}
\end{figure}

Figure~\ref{confused_trucks_MDPI} shows the tests of the quantum-tunnelling model trained on images of civilian and military trucks, with the results presented in the same style used to assess the model’s performance for the optical illusions of the Necker cube and Rubin's vase. In addition, the results presented in the bottom row panels of this figure can be used to simulate the classical perception of the images, i.e., a perception restricted to two discrete states of the cube~\cite{Mak24_APL}. To this end, I define the classical percept as a binary variable such that $\mathrm{Percept} = 1$ when the quantum probability $P_{\ket{1}} \geq 0.5$, and $\mathrm{Percept} = 0$ when $P_{\ket{1}} < 0.5$.

As can be seen in the two leftmost panels of  Fig.~\ref{confused_trucks_MDPI}, the model presented with an image of a civil truck predicts a higher probability of `civil' since the black output points in the bottom panel accept the value 1 more often than 0. However, as shown in Fig.~\ref{confused_trucks_MDPI},~right, the same model presented with an image of a military truck predicts a higher probability of ‘military’ since in this case the black output points in the right bottom panel accept the value 0 more often than 1.

Overall, the probability time dependencies shown in Fig.~\ref{confused_trucks_MDPI} follow patterns broadly consistent with those observed for the Necker cube (Fig.~\ref{Fig_Necker_result}) and Rubin’s vase (Fig.~\ref{Fig_Rubin_result}). However, closer inspection indicates that these dynamics exhibit greater confidence in their oscillations between 0 and 1, with a reduced prevalence of intermediate superposition states compared with the model outputs for the optical illusions. This effect is less pronounced for military trucks than for civilian ones.

On this basis, it can be inferred that military trucks are perceived by the neural network model as more illusion-like. This interpretation is reasonable, as military trucks, within the comparative framework adopted in this paper, are based on the same chassis as their civilian counterparts but are augmented with visually complex and potentially ambiguous features, such as protective grids, antennas and missile mounts. 

\begin{figure}
 \centering
 \includegraphics[width=0.7\textwidth]{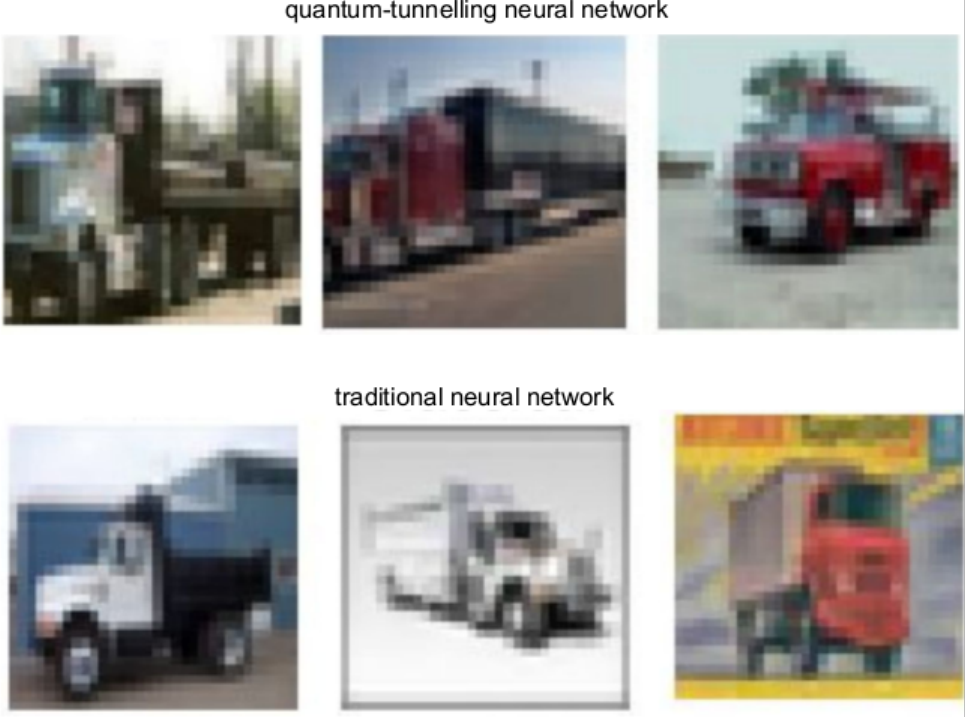}
 \caption{Representative examples of civilian trucks misclassified as military by the quantum-tunnelling~(top) and traditional~(bottom) neural network models. The discussion in the main text argues that, with the exception of the rightmost images in each row---rendered in red and visually reminiscent of fire trucks, whose distinctive features may be confounded with those of military vehicles producing specific read flames during the firing of missiles and projectiles---the classifications produced by the traditional neural network are clearly incorrect. By contrast, the classifications generated by the quantum-tunnelling network can be plausibly justified. In particular, some of the trucks identified as military exhibit colour schemes and cargo contours that could reasonably support such a classification.}
\label{confused_trucks_MDPI_2}
\end{figure}

Although the following discussion requires increased statistical significance to support fully conclusive claims, as shown in Fig.~\ref{confused_trucks_MDPI_2}, the images of the civilian trucks misclassified by the quantum-tunnelling neural network as military ones appear more consistent with the types of errors a `reasonable' human observer would make, compared with those produced by a traditional neural network. In particular, the conventional network tends to misclassify relatively obvious examples of civilian trucks, for instance those with clearly identifiable white cabins. (Here, I do not consider cases in which civilian vehicles are repurposed for combat use, as such scenarios are beyond the scope of the present system and are unlikely to be reliably addressed by any near-term technology.)

By contrast, the quantum-tunnelling network tends to misclassify as military those trucks that, although a reasonable human observer might still label as civilian, exhibit features that could plausibly suggest military use, such as darker cabins and semitrailer configurations towing large, green, seemingly elongated (e.g.~cylindrical, missile-like) structures. While the current dataset is not sufficiently large to support definitive conclusions, this correspondence is noteworthy and has been observed consistently across additional tests.

Interestingly, in Fig.~\ref{confused_trucks_MDPI_2} the right-most images misclassified by both traditional and quantum-tunnelling neural networks depict a fire truck and a truck with a red cabin. Fire trucks can, in practice, be mistaken for regular trucks or even military vehicles when their distinctive features, such as ladders, sirens or characteristic markings, are not clearly resolved or are misinterpreted. Psychological factors are also relevant here, particularly as red is widely associated with danger or aggression in human perception~\cite{Sol95, Kun15}.

It is also noteworthy that some training images of military vehicles in the investigated customised dataset include self-propelled artillery or combat tanks in action, producing a prominent red flame~\cite{Maks25_2}. Exposure to such imagery may bias the model towards a human-like association between red and military threat, thereby increasing the likelihood that fire trucks and other red-coloured vehicles are misclassified as military assets.

\section{Conclusions}
\subsection{Outlook}
Thus, beginning with well-documented behavioural paradoxes in human decision-making under risk, particularly in lottery and financial contexts, I have articulated the need for quantum cognition theory extended by machine learning techniques, thereby complementing the previous attempts to integrate quantum cognition with neural networks~\cite{Bus17}. While this theory has been established conceptually, it continues to seek firm grounding in practical applications. I have extended its scope to the perception of optical illusions and to social behaviour in networks, employing the mathematical formalism of a quantum oscillator.

Building on this foundation, I introduced a quantum barrier through which particles may tunnel, thereby advancing the model. This extension enables a more nuanced representation of ambiguous perception and facilitates the modelling of human mental states both in isolation, for example within social bubbles, and in interconnected settings such as social networks, where phenomena like opinion polarisation emerge.

By drawing parallels with other relevant theoretical approaches, including Khrennikov's social laser model and Aerts' interpretation of perception as a quantised phenomenon, I further extended the approach into the domain of machine learning. I demonstrated that a quantum-tunnelling representation of human beliefs and reasoning can serve as a fundamental data-processing component within neural network architectures. A deep neural network constructed on this basis exhibits a strong correspondence between the outputs of the quantum oscillator model and those of the network itself.

Subsequent evaluation on a customised dataset of real-world images of civil and military vehicles demonstrates the practical utility of the proposed approach across a range of realistic scenarios, thereby broadening the applicability of quantum cognition theory. Overall, the behaviour produced by the presented models is consistent with earlier theoretical predictions~\cite{Atm10, Bus12} and empirical observations~\cite{Gae98, Pia17, Joo20, Cho20}, supporting the interpretation of perceptual states as superpositions of $|0\rangle$ and $|1\rangle$.

These findings are also aligned with contemporary neuroscientific models, which characterise the perception of ambiguous figures as a continuous dynamical process influenced by stochastic fluctuations~\cite{Sto03, Mor07, Shp07, Cur08, Bue11, Ger12, Pan13, Ger14, Con15, Run16}. In several formulations, this dynamics is analogous to wave-like processes, in which phase evolution governs gradual transitions between perceptual states~\cite{Gla15}.

I also note the following possible extensions of the model that enhance its ability to capture complex behaviour and social phenomena.

\subsection{Quantum Randomisation of Neural Weights}
Here, I discuss an optional extension of the neural network algorithm outlined above. This approach allows for the introduction of experimental-physical quantum features into the model, making it potentially more suitable for tasks involving quantum dynamical systems exhibiting chaotic behaviour.

The physical processes underlying perceptual switching in ambiguous figures remain an open question~\cite{Leh95, Kor05}. A prominent class of models attributes such switching to nonlinear and chaotic dynamics~\cite{Kan89, Ino94, Sak95, Shi10, Che23}. More generally, the brain can be viewed as a complex dynamical system exhibiting nonlinear and chaotic behaviour across multiple scales~\cite{Bab88, McK94, Kor03}. It is therefore plausible that suitably constructed nonlinear physical models can approximate aspects of cognitive dynamics~\cite{Mak23_review}.

To introduce chaotic behaviour into the present model, I employ a quantum physical random number generator~\cite{Sym11, Haw15} to initialise the weight matrices $W^{(n)}$. Unlike pseudo-random generators~\cite{Rei92}, quantum generators produce intrinsically random sequences~\cite{Sym11, Haw15}. This ensures that the network remains unbiased with respect to the perceptual states of the Necker cube and avoids periodic repetition in its predictions~\cite{Her89, Fan18}. As a result, the network exhibits genuinely chaotic dynamics~\cite{Bru03}, consistent with earlier classical models~\cite{Kan89, Ino94}.

\subsection{Quantum Cognition for Human Sentiment Modelling with Language Models}
Drawing on the tutorial article~\cite{Maks25_1}, this brief note is intended to demonstrate that the application of quantum cognition theory in neural networks is not confined to machine vision, but extends more broadly to other machine learning architectures, including recurrent neural networks.

The original, traditionally-designed neural network model processes a dataset of words that a human would use to express their negative and positive sentiments (e.g., words one would use to provide a negative or positive feedback about purchased products or services). These words are structured as a two-column table, where the first column contains text phrases and the second assigns a binary label, positive or negative. The dataset is partitioned into two subsets:~a training set used for model fitting and a testing set used for performance evaluation.

While a model constructed in this manner is generally sufficient for standard applications such as chatbots, I extended its algorithm by integrating a quantum-tunnelling mechanism into both the learning and inference processes. I demonstrated that these modifications not only accelerate training~\cite{Maks25_1}, but also improve the model's capacity to handle ambiguous human verbal communication, potentially outperforming conventional approaches~\cite{Maks25_2}.
  
\vspace{6pt}

\section*{Author Contributions}
This is a sole-author paper and the author is fully accountable for its intellectual content.

\section*{Funding}
This research received no external funding.

\section*{Institutional Review}
Not applicable.

\section*{Informed Consent}
Not applicable.

\section*{Data Availability}
The source codes that implement the neural network models discussed in this paper are available at:
\begin{itemize}
\item \url{https://github.com/IvanMaksymov/Quantum-Tunnelling-Neural-Networks-Tutorial}
\item \url{https://github.com/IvanMaksymov/Cognition-in-Superposition/tree/main/Military-Trucks}
\item \url{https://github.com/IvanMaksymov/Cognition-in-Superposition}
\item \url{https://github.com/IvanMaksymov/OpinionPolarisation}
\end{itemize}

\section*{Acknowledgments}
The author acknowledges helpful discussions with Professor Ganna Pogrebna.

\section*{Conflicts of Interest}
The author declares no conflicts of interest.

\section*{Abbreviations}
\begin{tabular}{@{}ll}
CPT & cumulative prospect theory\\
CIFAR & Canadian Institute for Advanced Research database\\
EUT & expected utility theory\\
FDTD & finite-difference time-domain method\\
MNIST & Modified National Institute of Standards and Technology database\\
QT & quantum tunnelling\\
QZE & quantum Zeno effect\\
ReLU & rectified linear unit\\
\end{tabular}

\appendix
\section[\appendixname~\thesection]{Finite-Difference Time-Domain Method for Parabolic Barrier Modelling}
The Schr{\"o}dinger equation is a partial differential equation that governs the wavefunction of a quantum-mechanical system \cite{Gri04}. For a single electron that exists in a one-dimensional space it can be written as
\begin{equation}
  \label{eq:SE_fdtd}
  \imagi\hbar\frac{\partial \psi(x,t)}{\partial t}=\left[-\frac{\hbar^2}{2m}\frac{\partial^2}{\partial x^2} + V(x)\right]\psi(x, t)\,, 
\end{equation}
where $\psi(x, t)$ is a wave function, $\imagi$ is the imaginary unit, $m$ is the mass of the electron, $\hbar$ is Plank's constant and $V(x)$ is the potential that represents the environment where the electron exists.

I numerically solve Eq.~(\ref{eq:SE_fdtd}) using a finite-difference time-domain (FDTD) method\index{FDTD method} \cite{Sullivan}. I split the wave function $\psi(x, t)$ into the real and imaginary parts:
\begin{equation}
  \label{eq:Eq2_fdtd}
 \psi(x, t)=\psi_{re}(x, t)+{\imagi}\psi_{im}(x, t)\, 
\end{equation}
and rewrite Eq.~(\ref{eq:SE_fdtd}) as
\begin{gather}
  \label{eq:Eq3_fdtd}
  \frac{\partial \psi_{re}(x, t)}{\partial t} = -\frac{\hbar}{2m}\nabla^2 \psi_{im}(x, t)+\frac{1}{\hbar}V(x)\psi_{im}(x, t) \\ \nonumber
\frac{\partial \psi_{im}(x, t)}{\partial t} = \frac{\hbar}{2m}\nabla^2 \psi_{re}(x, t)-\frac{1}{\hbar}V(x)\psi_{re}(x, t)\,, 
\end{gather}
where $m\approx9.1093837\times10^{-31}$\,kg and $\hbar\approx1.054571817\times10^{-34}$\,J$\cdot$s. The coordinate $x$, time $t$ and wave function $\psi(x,t)$ are represented as discrete quantities using a spatially uniform mesh with the size $\Delta x$ and a temporal mesh with the size $\Delta t$. The $x$-coordinate becomes a vector of discrete elements $x_k = k \Delta x$, where $k = 1\dots N_x$ and $N_x$ is the number of nodes of the spatial mesh. Similarly, the discrete time instances are $t_n = n \Delta t$ with $n = 1\dots N_t$. The value of $\Delta t$ must be related to $\Delta x$ via the Courant stability criterion \cite{Sullivan}:
\begin{equation}
  \label{eq:Eq4}
  \Delta t = \frac{1}{8}\frac{2m}{\hbar} (\Delta x)^2\,. 
\end{equation}
Thus, I obtain a spatio-temporally discretised representation of Eq.~(\ref{eq:Eq3_fdtd}): 
\begin{gather}
  \label{eq:Eq5}
  \psi_{re}^{n}(k) = \psi_{re}^{n-1}(k)-\frac{1}{8}\left[\psi_{im}^{n-1/2}(k+1)-2\psi_{im}^{n-1/2}(k)+\psi_{im}^{n-1/2}(k-1)\right]+\frac{\Delta t}{\hbar}V(k)\psi_{im}^{n-1/2}(k)\\ \nonumber
  \psi_{im}^{n}(k) = \psi_{im}^{n-1}(k)+\frac{1}{8}\left[\psi_{re}^{n-1/2}(k+1)-2\psi_{re}^{n-1/2}(k)+\psi_{re}^{n-1/2}(k-1)\right]-\frac{\Delta t}{\hbar}V(k)\psi_{re}^{n-1/2}(k)\,. 
\end{gather}

The electron is modelled as a Gaussian-shape energy wave packet formed before the start of the simulation at the discrete instant of time $n=0$:
\begin{gather}
  \label{eq:Eq6}
  \psi_{re}^{0}(k) = \exp\left(-0.5\left(\frac{k-k_0}{\sigma}\right)^2\right) \cos\left(\frac{2\pi(k-k_0)}{\lambda}\right)\\ \nonumber
  \psi_{im}^{0}(k) = \exp\left(-0.5\left(\frac{k-k_0}{\sigma}\right)^2\right) \sin\left(\frac{2\pi(k-k_0)}{\lambda}\right)\,, 
\end{gather}
where $\lambda$ is the wavelength, $\sigma$ is the width of the Gaussian pulse and $k_0$ is the coordinate of origin of the pulse. Since the electron should be present somewhere in the potential well, the amplitudes of the wave functions are normalised as
\begin{equation}
  \label{eq:Eq7}
  \int_{-\infty}^{\infty} \psi^{*}(x)\psi(x) \,dx = 1\,. 
\end{equation}
The probabilities of funding the electron is the $|0\rangle$ and $|1\rangle$ regions of the potential well are calculated as 
\begin{gather}
  \label{eq:Eq8}
  P_{|0\rangle}=\int_{-\infty}^{x_{centre}} \psi^{*}(x)\psi(x) \,dx \\ 
  P_{|1\rangle}=\int_{x_{centre}}^{\infty} \psi^{*}(x)\psi(x) \,dx\,, 
\end{gather}
where $P_{|0\rangle}+P_{|1\rangle}=1$. The following physically meaningful model parameters were used in the simulations~\cite{Cha74}: $\Delta x=0.1\times10^{-11}$\,m, $N_x=4001$, $N_t=3\times10^6$, $\lambda=1.6\times10^{-10}$\,m and $\sigma=1.6\times10^{-10}$\,m.

\section[\appendixname~\thesection]{Finite-Difference Method for Social Network Modelling}
I solve the Schr{\"o}dinger equation that defines the eigenfunctions\index{eigenfunction} corresponding to the eigenvalues $E$ of the Hamiltonian operator $\hat{H}$, as \cite{Gri04}
\begin{equation}
  \label{eq:SE_ch2}
  \hat{H}\psi({\bf r}) \equiv \left[-\frac{\hbar^2}{2m}\Delta + V({\bf r})\right]\psi({\bf r}) = E\psi({\bf r})\,, 
\end{equation}
where $\hbar$ is Plank's constant, $\Delta$ is the Laplacian operator, $m$ is the mass of the electron and $V({\bf r})$ is the scalar potential. I employ a finite-difference method in which Eq.~(\ref{eq:SE_ch2}) is discretised along the coordinate $x$, such that $V(x)$ is represented as a vector of $N$ equally spaced points with a step size of $h_x$ \cite{Hal22}. Using a second-order central finite-difference scheme, I express
\begin{equation}
  \label{eq:deriv2}
  \psi''(x_i) \approx \frac{1}{h_x^2}\left[\psi(x_{i-1})-2\psi(x_{i})+\psi(x_{i+1})\right]\,. 
\end{equation}
Substituting Eq.~(\ref{eq:deriv2}) into Eq.~(\ref{eq:SE_ch2}) yields a matrix equation \cite{Hal22} 
\begin{equation}
  \label{eq:deriv3}
  -\frac{\hbar^2}{2mh_x^2}
    \left\lceil
    \begin{matrix}
    -2 & 1 & 0 & 0 & 0 \\
    1 & -2 & 1 & 0 & 0 \\
    0 & 1 & -2 & 1 & 0 \\
    ~ & ~ & ~ & \cdots & ~\\
    0 & 0 & 0 & 1 & -2        
    \end{matrix}
    \right\rceil
    \left\lceil
    \begin{matrix}
    \psi(x_1)\\
    \psi(x_2)\\
    \psi(x_3)\\
    \cdots\\
    \psi(x_N)        
    \end{matrix}
    \right\rceil
    +
    \left\lceil
    \begin{matrix}
    V(x_1)\psi(x_1)\\
    V(x_2)\psi(x_2)\\
    V(x_3)\psi(x_3)\\
    \cdots\\
    V(x_N)\psi(x_N)        
    \end{matrix}
    \right\rceil
    =
    E\left\lceil
    \begin{matrix}
    \psi(x_1)\\
    \psi(x_2)\\
    \psi(x_3)\\
    \cdots\\
    \psi(x_N)        
    \end{matrix}
    \right\rceil\,, 
\end{equation}
where $m\approx9.1093837\times10^{-31}$\,kg and $\hbar\approx1.054571817\times10^{-34}$\,J$\cdot$s. I also set $h_x=2 \times 10^{-10}$\,m. 

Using the virtual points $x_0$ and $x_{N+1}$, which do not directly participate in the calculation but assist in evaluating the neighbouring points, I apply Floquet periodic boundary conditions $\psi(x_0)=\psi(x_N)\exp(-\imagi ka)$ and $\psi(x_{N+1})=\psi(x_{1})\exp(\imagi ka)$ (see Ref.~\cite{Mak_thesis}, p.~82). Here, $\imagi$ denotes the unit imaginary number, $k$ represents the wavevector and $a$ is the period of repetition of the potential profile $V(x)$. For an isolated potential well, the boundary conditions reduce to $\psi(x_0)=\psi(x_{N+1})=0$. The introduction of Floquet periodic boundary conditions modifies the last element of the first row and the first element of the last column of the tridiagonal matrix on the left-hand side of Eq.~(\ref{eq:deriv3}).

The numerical solution of Eq.~(\ref{eq:deriv3}), obtained using Python’s or MATLAB's standard {\tt eigs} subroutine, produces the eigenvalues $E$ for each discrete wavevector $k$ within the first Brillouin zone. The energy-level diagrams, shown alongside the dispersion curves, are generated by integrating the $E$ values over all $k$ points included in the calculation. 

\section[\appendixname~\thesection]{Quantum-Tunnelling Activation Function}
In quantum mechanics~\cite{Gri04}, the rectangular potential barrier serves as a canonical example of quantum tunnelling, where the one-dimensional, time-independent Schr{\"o}dinger equation governs the behaviour of an electron approaching the barrier\index{energy!barrier}~\cite{McQ97}.
\begin{equation}
  \label{eq:SE_2}
  \left[-\frac{\hbar^2}{2m}\frac{d^2}{d x^2} + V(x)\right]\psi(x) = E\psi(x)\,, 
\end{equation}
where $\psi(x)$ is a wave function, $m\approx9.1093837\times10^{-31}$\,kg is the mass of the electron, $\hbar\approx1.054571817\times10^{-34}$\,J$\cdot$s is Plank's constant and $E$ is the energy of the electron. The profile of the potential barrier  is 
\begin{equation}\label{eq:Vx}
    V(x) = 
    \begin{cases}
      0 & \text{for }\,x < 0\\
      V_0 & \text{for }\,0 < x \le a\\
      0 & \text{for }\,x > a\,.
    \end{cases}       
\end{equation}

Omitting the auxiliary derivations~\cite{McQ97}, the electron tunnelling behaviour can be quantified by computing the transmission coefficient from the solution of Eq.~(\ref{eq:SE_2}) for the potential barrier given by Eq.~(\ref{eq:Vx}). The solution of the Schr{\"o}dinger equation can be expressed as a superposition of left- and right-moving waves.
\begin{equation}
  V(x) =
    \begin{cases}
      \psi_L(x)=A_1e^{\imagi kx}+A_2e^{-\imagi kx}, & x < 0\\
      \psi_C(x)=B_1e^{\imagi\kappa x}+B_2e^{-\imagi\kappa x}, & 0 < x \le a\\
      \psi_R(x)=C_1e^{\imagi kx}+C_2e^{-\imagi kx}, & x > a\,,
    \end{cases}       
\end{equation}
where $\imagi$ is the imaginary unit, $k=\sqrt{2mE/\hbar^2}$ and $\kappa=\sqrt{2m\alpha/\hbar^2}$ with $\alpha=E-V_0$ (the special cases $E=0$ and $E=V_0$ are treated separately). The coefficients $A, B, C$ can be found from the boundary conditions at $x=0$ and $x=a$, where $\psi(x)$ and its derivative must be continuous. 

For $E<V_0$, there is a non-zero transmission probability   
\begin{equation}
  \label{eq:eq1}
  T\vert_{E<V_0} = \left(1-\beta\sinh^2(\kappa_1 a)\right)^{-1}\,, 
\end{equation}
where $\beta=\dfrac{V_0^2}{4E\alpha}$ and $\kappa_1=\sqrt{-2m\alpha/\hbar^2}$. For $E>V_0$
\begin{equation}
  \label{eq:eq2}
  T\vert_{E>V_0} = \left(1+\beta\sin^2(\kappa a)\right)^{-1}\,, 
\end{equation}Finally, the expression for $E=V_0$ is obtained by taking the limit of $T$ as $E$ approaches $V_0$, resulting in
\begin{equation}
  \label{eq:eq3}
  T\vert_{E=V_0} = \left(1+\frac{ma^2V_0}{2\hbar^2}\right)^{-1}\,. 
\end{equation}

Differentiating Eqs.~(\ref{eq:eq1}--\ref{eq:eq2}) with respect to $E$, I obtain the expression needed for the backpropagation part of the neural network algorithm
\begin{equation}
\begin{gathered}\label{eq:diff1}
  T^\prime\vert_{E<V_0}=-\beta\biggl(\frac{\sinh^2(\delta_1)}{E}+ \\\frac{\sinh^2(\delta_1)-\delta_1\cosh(\delta_1)\sinh(\delta_1)}{\alpha}\biggr)T^2\vert_{E<V_0}\,, \\
  T^\prime\vert_{E>V_0}=\beta\biggl(\frac{\sin^2(\delta)}{E}+\frac{\sin^2(\delta)-\delta\cos(\delta)\sin(\delta)}{\alpha}\biggr)T^2\vert_{E>V_0}\,, \\
  T^\prime\vert_{E=V_0}=\frac{4V_0a^4m^2+6a^2\hbar^2m}{3V_0^2a^4m^2+12V_0a^2\hbar^2m+12\hbar^4}\,,
\end{gathered}
\end{equation}
where $\delta=\kappa a$ and $\delta_1=\kappa_1a$.

\section[\appendixname~\thesection]{Neural Network Perception Rubin’s vase}
\begin{figure}[t]
\centering
 \includegraphics[width=0.8\textwidth]{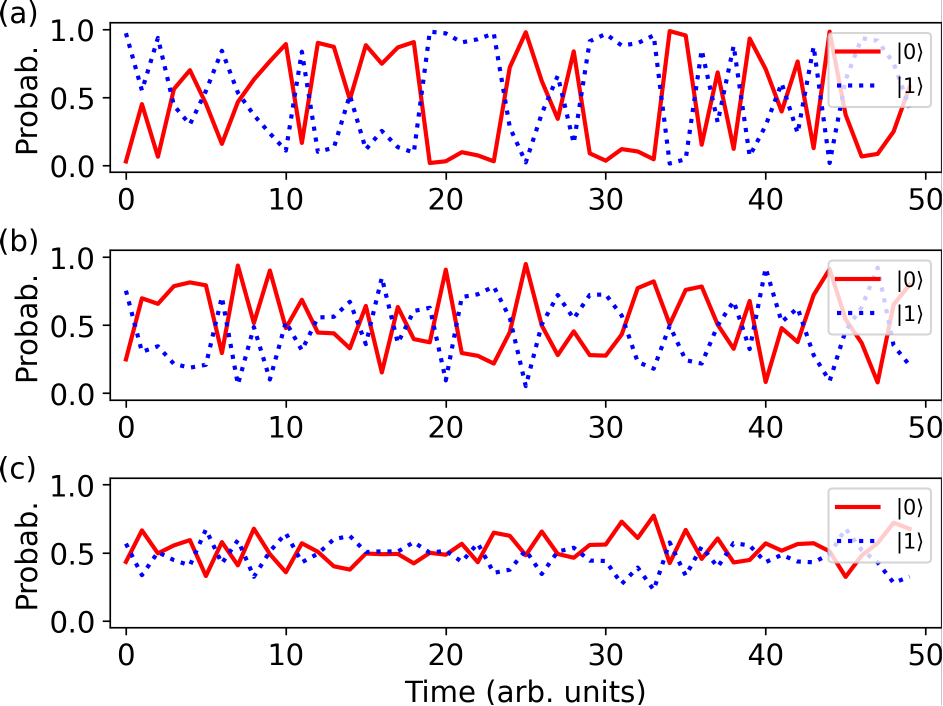}
 \caption{Perceptual switching curves produced by the quantum-tunnelling neural network trained on Rubin’s vase. The data points with probability $P_{|0\rangle}=0$ or $P_{|1\rangle}=1$ correspond to the fundamental perceptual states of Rubin’s vase. The remaining data points are in a superposition of states $|0\rangle$ and $|1\rangle$ with $P_{|0\rangle}+P_{|1\rangle}=1$.\label{Fig_Rubin_result}}
\end{figure}

As shown in Fig.~\ref{Fig_Rubin_result}, perceptual switching for Rubin’s vase follows a similar qualitative pattern to that of the Necker cube, discussed in the main text. However, the neural network achieves clearer separation between the two perceptual states, even for thick barriers (Fig.~\ref{Fig_Rubin_result}c), without requiring an increased number of training epochs. This observation is consistent with the hypothesis that Rubin’s vase combines figure–ground segmentation with bistable perception \cite{Kha21_1}. In particular, observers may interpret the image not only as faces versus vase, but also as a light object on a dark background (or vice versa), providing an additional reference frame that facilitates disambiguation.

\section[\appendixname~\thesection]{Discussion of Connections with Other Quantum Cognition Models}
The quantum Zeno effect (QZE) has been widely employed to model bistable perception~\cite{Atm04, Atm10, Atm13, yearsley2016zeno, Kor17}, providing a complementary perspective to the quantum tunnelling models developed in this paper. The QZE refers to the inhibition of the evolution of a quantum system under frequent observation, effectively stabilising its state~\cite{Mis77, Harr17}. In other words, measurement does not merely reveal the state of a system but actively constrains its dynamics.

This principle has been extensively studied in quantum optics and quantum information science, where repeated measurement can suppress transitions such as quantum tunnelling. Experimental studies have demonstrated that increasing the frequency of observation reduces the probability of tunnelling events, highlighting a direct interplay between measurement and dynamical evolution. 

Within the context of human cognition, the QZE provides a natural mechanism for understanding how sustained attention or repeated observation can stabilise mental states~\cite{yearsley2016zeno}. Behaviour and perception may be viewed as comprising a range of potential states, with only a subset becoming manifest under specific conditions. Frequent observation, whether external or self-imposed, may constrain this range, effectively `freezing' the system in a particular configuration. 

A similar argument applies to the perception of ambiguous figures. Empirical and theoretical studies indicate that deliberate modulation of attention, including factors such as eye-blink frequency or sustained focus, can alter perceptual switching dynamics~\cite{Kor05, Kor17}. From a QZE perspective, increased observational frequency suppresses transitions between competing interpretations, while reduced attention allows more frequent alternation.

In contrast, the quantum tunnelling model employed in this paper emphasises transitions between states via barrier penetration, capturing the probabilistic switching behaviour observed in perception and decision-making. The QZE and tunnelling frameworks are therefore not mutually exclusive but describe complementary regimes of the same underlying dynamics:~the former accounts for stabilisation under frequent observation, while the latter explains spontaneous transitions driven by intrinsic system dynamics.

Finally, it is important to note that other quantum-mechanical concepts, including interference effects, violations of classical probability structures and information-theoretic constraints~(for a review see, e.g., Ref.~\cite{Pot22}),  also contribute to a more complete description of cognition. While these aspects are not explicitly incorporated into the present model, they point to a broader theoretical landscape in which quantum-inspired approaches can offer valuable insights.

\bibliographystyle{unsrt}  
\bibliography{refs}

\end{document}